\documentstyle[emulateapj,onecolfloat]{article}

\begin{document}
\twocolumn[
\title{A search for the most massive galaxies: Double Trouble?}
\author{M. Bernardi$^{1}$, R. K. Sheth$^1$, R. C. Nichol$^2$, 
C. J. Miller$^3$, D. Schlegel$^4$, J. Frieman$^{5,6}$, 
D. P. Schneider$^7$, M. Subbarao$^5$,  D. G. York$^5$, J. Brinkmann$^8$}

\begin{abstract}
We describe the results of a search for galaxies with large 
($\ge 350$~kms$^{-1}$) velocity dispersions.  
The largest systems we have found appear to be the extremes of the 
early-type galaxy population:  compared to other galaxies with similar 
luminosities, they have the largest velocity dispersions and the 
smallest sizes.  However, they are not distant outliers from the 
Fundamental Plane and mass-to-light scaling relations defined by 
the bulk of the early-type galaxy population.  
They may host the most massive black holes in the Universe, and 
their abundance and properties can be used to constrain galaxy formation 
models.
Clear outliers from the scaling relations tend to be objects in 
superposition (angular separations smaller than 1~arcsec), 
evidence for which comes sometimes from the spectra, sometimes from 
the images, and sometimes from both.  
The statistical properties of the superposed pairs, e.g., the 
distribution of pair separations and velocity dispersions, can be used 
to provide useful information about the expected distribution of image 
multiplicities, separations and flux ratios due to gravitational lensing 
by multiple lenses, and may also constrain models of their interaction 
rates.  
\end{abstract}
\keywords{galaxies: elliptical --- galaxies: evolution --- 
          galaxies: fundamental parameters --- galaxies: photometry --- 
          galaxies: stellar content}
]

\footnotetext[1] {Department of Physics and Astronomy, 
                  University of Pennsylvania, Philadelphia, PA 19104}
\footnotetext[2] {Institute of Cosmology and Gravitation (ICG), Mercantile House, Hampshire Terrace, University of Portsmouth, Portsmouth, PO1 2EG, UK}
\footnotetext[3] {Cerro-Tololo Inter-American Observatory, NOAO, Casilla 603, 
                  La Serena, Chile}
\footnotetext[4] {Princeton University Observatory, Princeton, NJ 08544}
\footnotetext[5] {University of Chicago, Astronomy \& Astrophysics Center, 
                  5640 S. Ellis Ave., Chicago, IL 60637}
\footnotetext[6] {Fermi National Accelerator Laboratory, P.O. Box 500,
Batavia, IL 60510}
\footnotetext[7] {Department of Astronomy and Astrophysics, The Pennsylvania State University, University Park, PA 16802}
\footnotetext[8] {Apache Point Observatory, 2001 Apache Point Road, P.O. Box 59, Sunspot, NM 88349-0059}

\section{Introduction}
Giant early-type galaxies are expected to be more massive than spirals.  
They typically have line-of-sight velocity dispersions larger than 
200~km~s$^{-1}$.  The massive cD galaxies at the centers of some groups 
and clusters are expected to be substantially more massive, and are 
thought to be amongst the most massive galaxies in the universe 
(Dressler 1979; Porter, Schneider \& Hoessel 1984).  
Published measurements of velocity dispersions of cDs tend to not 
exceed $\sim 400$~km~s$^{-1}$ (for reference, the line-of-sight 
velocity dispersion of a dark matter halo with mass 
$5\times 10^{13}\,h_{100}^{-1}M_\odot$ at $z = 0.2$ is 
$\sim 400$~km~s$^{-1}$), but it is not clear whether this reflects 
a bona fide physical upper-limit, or if it is simply that such objects 
are rare, and surveys to date have not probed sufficiently large volumes 
to find them.  Indeed, extrapolation of the distribution of early-type 
galaxy velocity dispersions suggests that the abundance of objects with 
velocity dispersions in excess of 350~km~s$^{-1}$ and 400~km~s$^{-1}$ 
should be $4\times 10^{-7}~(h_{70}^{-1}$Mpc)$^{-3}$ 
and $1.6\times 10^{-8}~(h_{70}^{-1}$Mpc)$^{-3}$, respectively 
(Sheth et al. 2003).  The Sloan Digital Sky Survey (hereafter SDSS; 
York et al. 2000) is just beginning to probe a sufficiently large 
volume that such systems, if they exist, should appear in significant 
numbers.  The SDSS First Data Release covers an area of approximately 
2000 square degrees (Abazajian et al. 2003).  In a spatially flat 
cosmological model with $\Omega_0=0.3$ and $H_0=70$~km~s$^{-1}$~Mpc$^{-1}$, 
which we adopt in what follows, the comoving volume of a cone 2000 
square degrees on the sky out to $z=0.3$ is $3.34\times 10^8$~Mpc$^3$.  

In what follows, we describe a search for systems with extreme velocity 
dispersions in the SDSS.  
We select a sample of early-type galaxies
from the SDSS survey following techniques described by 
Bernardi et al. (2003a).  This selection is described in 
Section~\ref{sample}.  The SDSS spectroscopic pipeline reports that 
about 100 of these objects have velocity dispersions in excess of 
350~km~s$^{-1}$.  

Section~\ref{unusual} presents the results of a reanalysis of the 
images and spectra of these objects.  Although all these objects were 
classified as single galaxies by the SDSS photometric pipeline, almost 
half have spectra and/or images which indicate that they are, in fact, 
superpositions.  Evidence for superposition also comes from consideration 
of the location of these objects relative to the early-type galaxy 
scaling relations such as the Fundamental Plane, the mass-to-light ratio, 
and the correlation between color and velocity dispersion.  
An Appendix provides estimates of the likelihood of projection, and 
describes how we estimate the velocity dispersions and separations of 
the individual components.  These are interesting objects in their own 
right.  

The other $\sim 70$ objects with estimated velocity dispersions in 
excess of 350~km~s$^{-1}$ are not obviously superpositions.  Although 
it is difficult to argue conclusively that they really are single 
galaxies, Section~\ref{mg2test} uses the Mg$_2$-$\sigma$ correlation 
to argue that at least some of these objects are extremely likely to 
be singles.  Some of these objects are in crowded fields, whereas others 
are quite isolated.  They appear to be at the extreme tails of the 
early-type galaxy scaling relations, but they are not obvious outliers.  
These are interesting objects for follow-up study.  


We discuss some implications of our findings in Section~\ref{discuss}.  
For instance, the single galaxies with the largest velocity dispersions 
in our sample potentially host the most massive black holes in the 
Universe; 
existence of objects with large velocity dispersions constrains models 
of the gas cooling and baryonic contraction associated with galaxy 
formation in dark matter halos;  
comparison of the predicted number of superpositions with the number we 
think we have seen can be used to constrain the amount of extinction due 
to dust in early-type galaxies and/or the density profiles of clusters 
on very small scales;  and 
our sample of close pairs is useful for models of gravitational lensing 
by binary- or more complex lenses.

\section{The sample}\label{sample}
All the objects we analyze were selected from the 
Sloan Digital Sky Survey (SDSS) database.
See York et al. (2000) for a technical summary of the SDSS project; 
Stoughton et al. (2002) for a description of the Early Data Release; 
Abazajian et al. (2003) for a description of DR1, the First Data Release; 
Gunn et al. (1998) for details about the camera; 
Fukugita et al. (1996), Hogg et al. (2001) and Smith et al. (2002) 
for details of the photometric system and calibration; 
Lupton et al. (2001) for a discussion of the photometric data reduction 
pipeline; 
Pier et al. (2002) for the astrometric calibrations; 
Blanton et al. (2003) and Strauss et al. (2002) for details of the 
tiling algorithm and target selection.  

We selected all objects targeted as galaxies and with Petrosian 
apparent magnitude $14.5 \le r_{\rm Pet}\le 17.75$.  
To extract a sample of early-type galaxies we then chose the subset 
with the spectroscopic parameter {\tt eclass < 0} ({\tt eclass} classifies 
the spectral type based on a Principal Component Analysis), 
and the photometric parameter {\tt fracDev$_r$ > 0.8}. 
(The parameter {\tt fracDev} is 
a seeing-corrected indicator of morphology.  It is obtained by taking 
the best fit exponential and de Vaucouleurs fits to the surface 
brightness profile, finding the linear combination of the two that 
best-fits the image, and storing the fraction contributed by the 
de Vaucouleurs fit.) We removed galaxies with problems 
in the spectra (using the {\tt zStatus} and {\tt zWarning} flags).
From this subsample, we finally chose those objects for which the 
spectroscopic pipeline had measured velocity dispersions (meaning 
that the signal-to-noise ratio in pixels between the restframe 
wavelengths 4200\AA\ and 5800\AA\ is S/N $>10$).  This gave a sample 
of 39320 objects, with photometric parameters output by version 
${\tt V5.4}$ of the SDSS photometric pipeline and ${\tt V.23}$ 
reductions of the spectroscopic pipeline.  

We considered increasing the volume of our sample to $z=0.4$, by also 
including those objects which the SDSS targets as Luminous Red Galaxies 
(Eisenstein et al. 2001).  Most of the luminous objects in our main 
early-type galaxy sample are, in fact, also LRGs; so the main effect 
of including the LRG sample is to reduce the magnitude limit for the 
reddest objects.  However, many of the fainter LRGs which are not 
already in our main sample tend to have spectra with small S/N ratios, 
making it difficult to assign reliable velocity dispersions (indeed, 
the SDSS pipeline estimates velocity dispersions only if S/N $>10$).  
We have checked that the LRGs with larger S/N ratios follow the same 
scaling relations as the main sample, so we decided to use only galaxies 
drawn from the main sample, and to not include objects which were 
targeted specifically as LRGs.  

\begin{figure}[b]
 \centering
 \epsfxsize=\hsize\epsffile{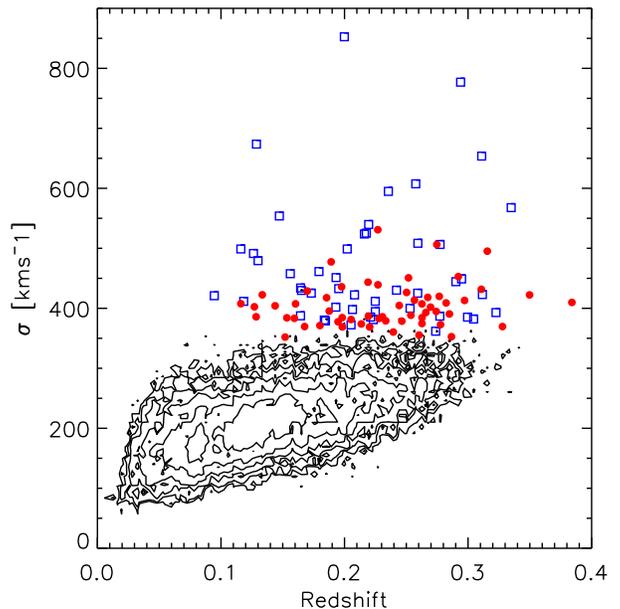}
 \caption{Velocity dispersions of early-type galaxies as a function of 
          redshift.  Contours, spaced in factors of two from the maximum, 
          represent the distribution of objects in the main sample, 
          blue squares show the objects we think are superpositions, 
          and red circles show objects for which the evidence 
          for superposition is weakest.  }
 \label{vz}
\end{figure}

The contours in Figure~\ref{vz} show the distribution of velocity 
dispersions in our sample.  (The convention is to report velocity 
dispersions which have been corrected to an aperture that is 1/8 times 
the half-light radius (e.g. J{\o}rgensen, Franx \& Kj{\ae}rgaard 1995).
We follow this convention, so the aperture 
corrected velocity dispersions reported below slightly larger than 
the measured values.)
Because luminosity and velocity dispersion are correlated 
(Bernardi et al. 2003b show that $L\propto \sigma^4$)  
the objects with small velocity dispersions are likely to fall below the 
magnitude limit of our sample at higher redshifts.  This accounts for 
the weak trend with redshift.  


Of the main sample of objects represented by the contours, the 
spectroscopic pipeline reports that 105 have velocity dispersions 
$\sigma>350$~km~s$^{-1}$.  These estimates assume that the spectrum 
really is that of a single object.  In Appendix~\ref{like}, we argue 
that the likelihood of a superposition which is sufficiently close that 
the photometry treats the blend as a single object is about one in every 
three hundred of the objects selected as early-types, so that some of 
the most distant luminous objects are likely to be superpositions.  
Therefore, we have performed our own estimates of the velocity dispersion 
for all these objects.  
Our analysis is described in some detail in Appendix~\ref{evidence}, 
where we conclude that most of the objects with $\sigma\ga 500$~km~s$^{-1}$ 
are actually superpositions.  
The small blue squares in Figure~\ref{vz} show objects we identified 
as superpositions, 
and the small red circles show objects for which the evidence for 
superposition is less compelling.  

Tables~\ref{tab:singles} and~\ref{tab:doubles} list some of the important 
measured parameters of these objects.  For objects which are not obvious 
superpositions Table~\ref{tab:singles} provides the object name, redshift, 
absolute $r-$band magnitude, restframe {\tt model} $g-r$ color, 
$r-$band size, aperture corrected velocity dispersion and associated 
measurement errors.  
The magnitudes and sizes we use are those which come from the SDSS 
pipeline fits of a deVaucoleur's profile to the surface-brightness 
distribution.  
In addition, the Table reports the S/N ratio of the spectrum, 
and for objects with S/N $\ge 18$, it reports the Mg$_2$ line-strength 
(estimates at lower S/N are very unreliable).  
These values are used in Section~\ref{mg2test}, which describes the 
results of an additional test of superposition.  
An asterix has been placed after the S/N values of all objects for which 
the evidence for superposition is weakest.  These objects show little 
irregularity in imaging and little asymmetry in the cross-correlation 
function---they may well be single objects.  

Table~\ref{tab:doubles} provides analogous information for the objects 
which are almost certainly superpositions.  For these objects, the 
measured parameters of the blend are almost certainly not those of the 
individual components, so we have chosen to not report the measurement 
errors (which are similar in magnitude to those in~Table~\ref{tab:singles}).  
However, we have included an estimate of the line-of-sight separation 
(in km~s$^{-1}$) between the two components, obtained from the analysis 
described in Appendices~\ref{evidence} and~\ref{veldisps}.  

Figure~\ref{uniqueimgspec} shows fields which are a few arcseconds 
on a side (for reference, the angular diameter distance 
corresponding to one arcsecond at $z=0.3$ is $4.4h_{70}^{-1}$kpc), 
and slightly larger fields,  $\sim 1$\arcmin\ on a side, centered on 
each object which we did not classify as a superposition.   
(This larger scale is close to the minimum spacing between SDSS fibers, 
55\arcsec, so, typically, only the central object in the field will 
have an SDSS spectrum, even if others satisfied the magnitude limit.)  
The figure also shows the result of two 
different techniques for estimating the velocity dispersion, as well 
as sections of the spectra of these objects, with the best-fitting 
template spectrum superimposed.  In these figures, and these figures 
only, we show the measured velocity dispersion, before correcting to 
an aperture of $r_e/8$, since it is these measured values which are 
altered by superposition.  
Clearly, some of our objects are in relatively crowded fields, whereas 
others are rather isolated.  A similar analysis of the objects 
classified as superpositions is shown in Figure~\ref{doubleimgspec}.  
(The electronic version of this article shows similar figures for all 
the objects in our sample; only a few representative examples are 
presented here.)

\section{Normal or Anomalous?}\label{unusual}
We have checked if the objects which are not obvious superpositions 
are a distinct population.  
Evidence that they are not substantially different from the bulk of 
early-type galaxies is presented in Figure~\ref{scalings}.   
In each panel, contours represent the full early-type galaxy sample, 
blue squares represent the objects we identified as superpositions, 
and red circles represent objects which could be either singles or 
doubles.  
Error bars indicate the uncertainty in the measurements of the 
velocity dispersions, sizes, masses, and colors (in some cases they 
are smaller than the dimensions of the symbols; for clarity, and to 
illustrate the magnitude of the typical color error, we have only 
shown errors for some of the objects).  
The solid line in the top left panel shows the Fundamental Plane 
relation reported by Bernardi et al. (2003c); although this fit is 
based on old photometric and spectroscopic reductions, it's slope 
provides a good description of the new data. 
Solid lines in the other panels show fits to the other scaling 
relations derived for the full early-type sample.  
Following Bernardi et al. (2003b,c), the luminosities and colors have 
been corrected for evolution (the correction is $0.85z$ to 
$M_{r}$ and $0.3z$ to $M_{g}-M_{r}$), and the fits correct for 
the effect of the magnitude limit of the SDSS.

\begin{figure}
 \epsfxsize=0.78\hsize\epsffile{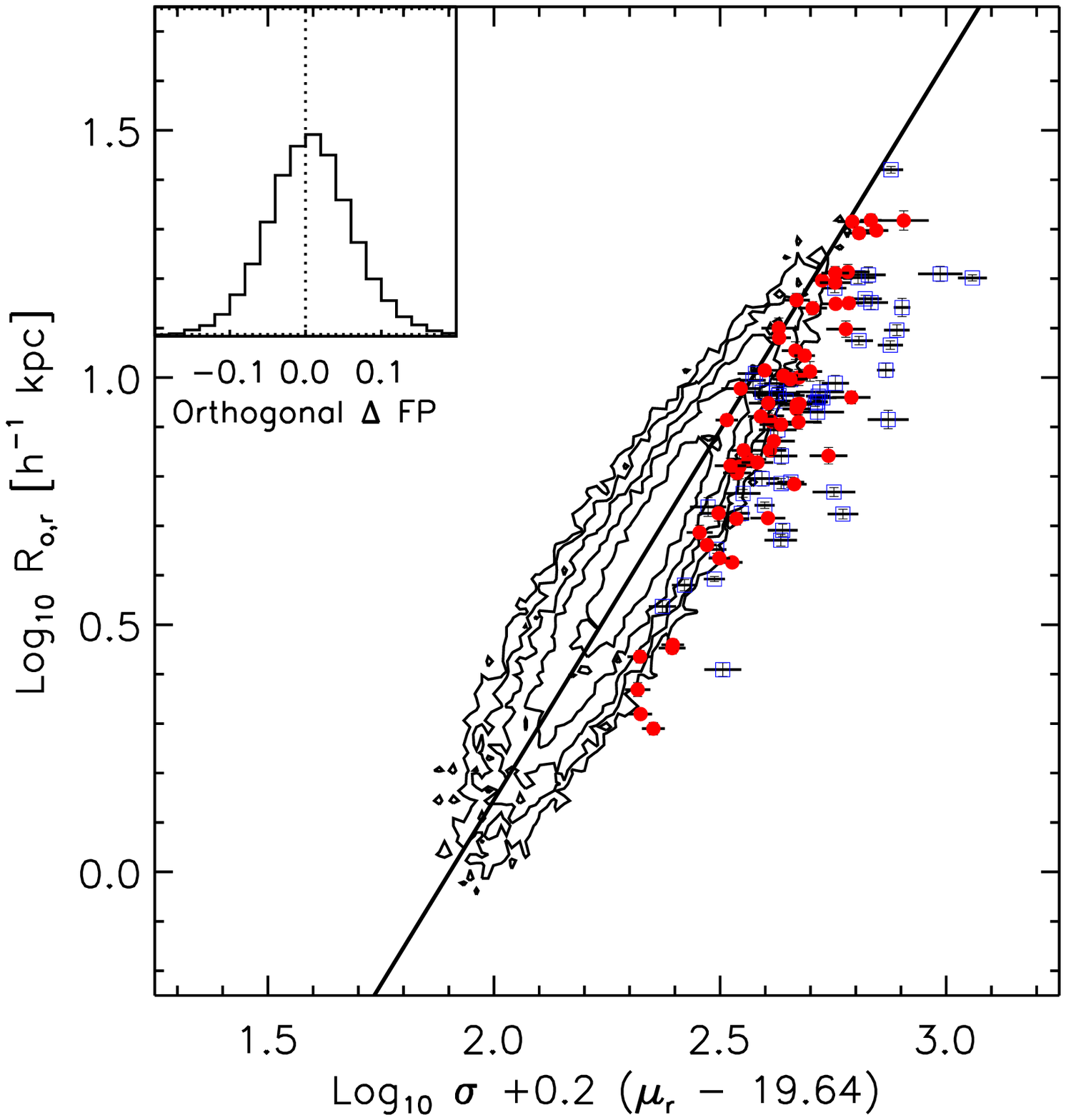}
 \epsfxsize=0.78\hsize\epsffile{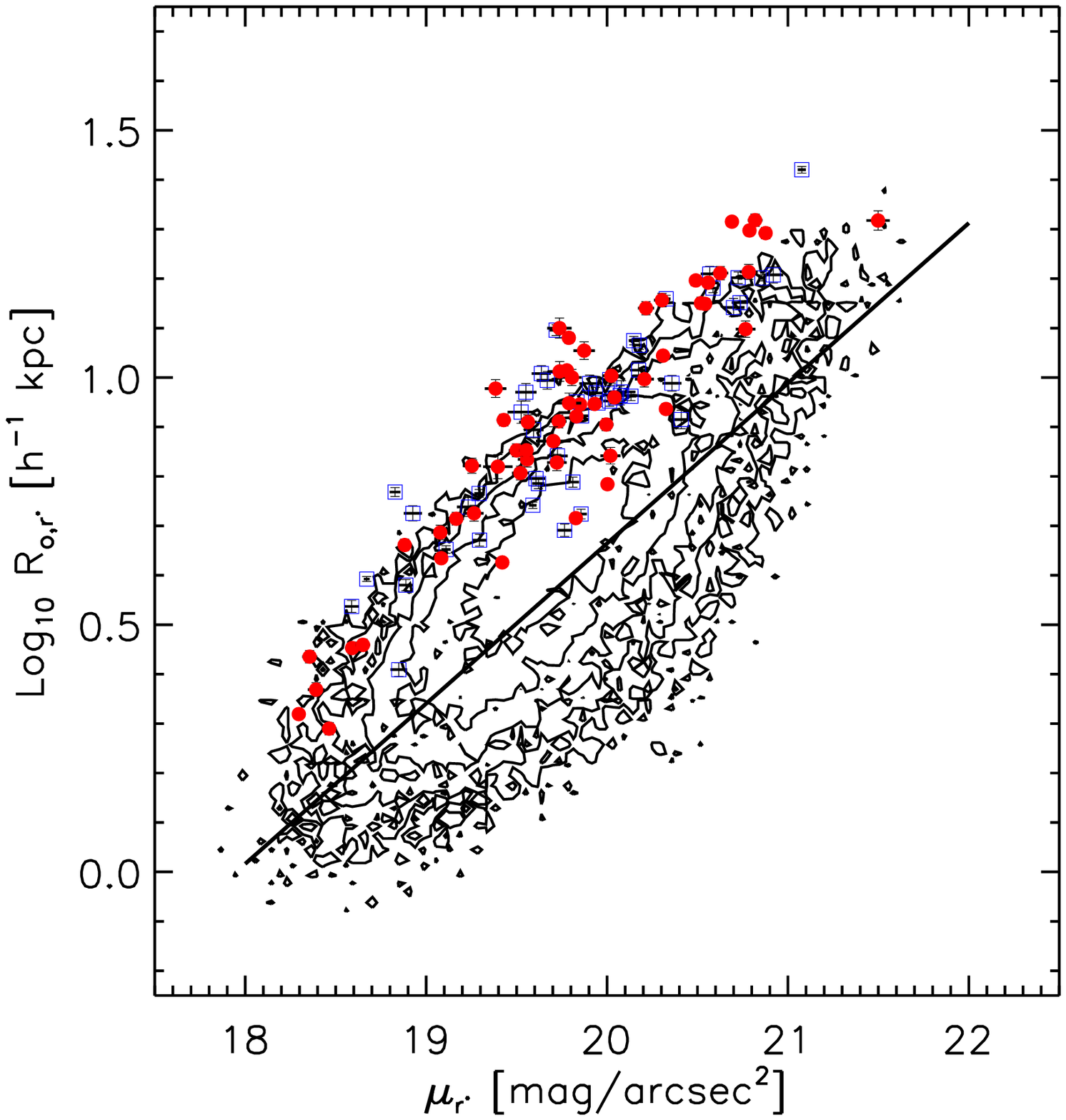}
 \epsfxsize=0.78\hsize\epsffile{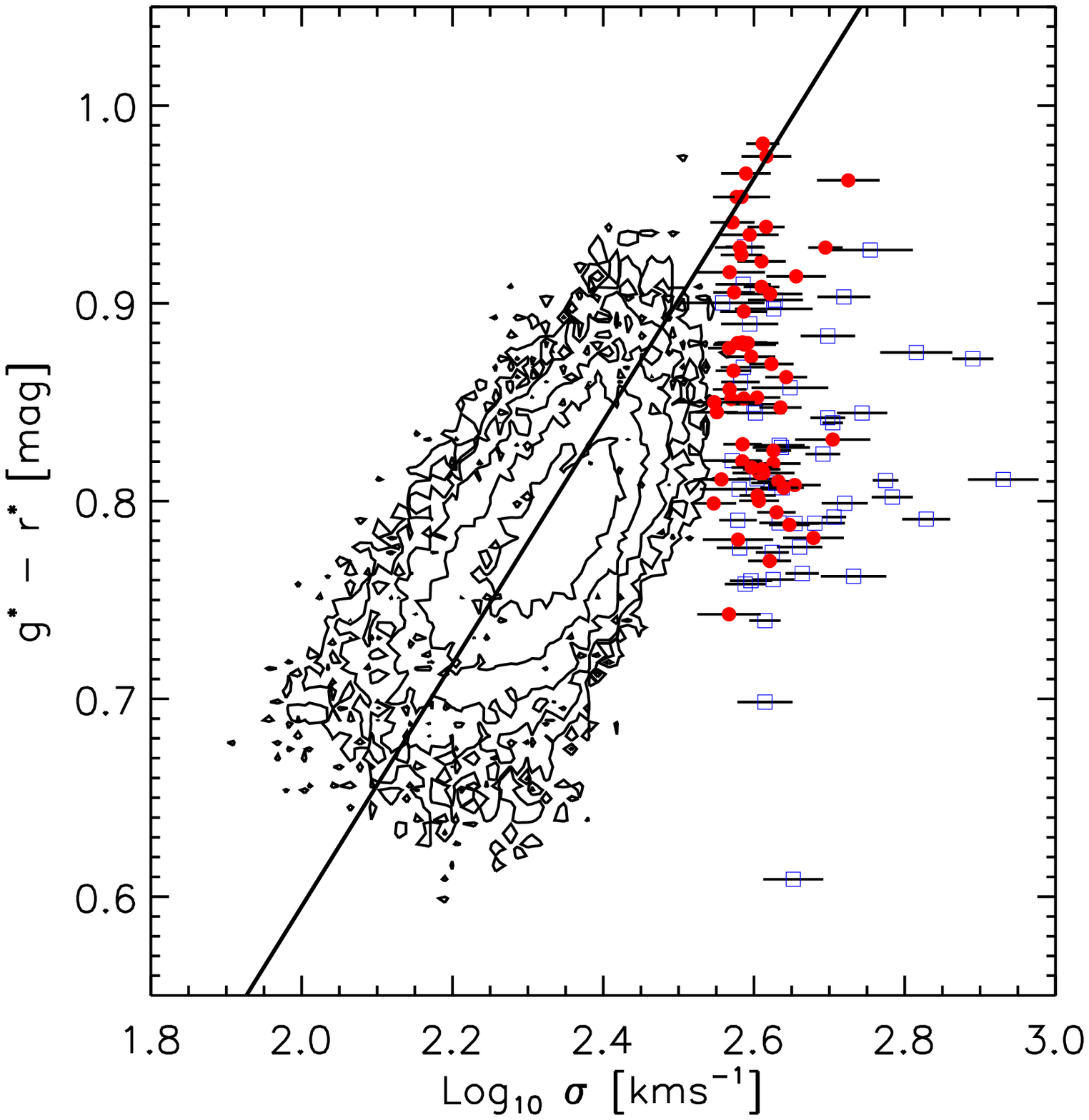}

 \caption{Location of our sample of objects (red circles) with respect 
          to the Fundamental Plane, the mass-luminosity relation, the 
          size-surface brightness relation, the size-luminosity 
          relation, the color-$\sigma$ and luminosity-$\sigma$ 
          relations. Luminosities and colors have been corrected for 
          evolution (following Bernardi et al. 2003b).  
          Velocity dispersions have been corrected for aperture effects.  
          Contours represent normal early-type galaxies, 
          blue squares represent objects we are 
          quite sure are superpositions, and red circles represent objects 
          which could be either singles or doubles.}
 \label{scalings}
\end{figure}

\begin{figure}
 \epsfxsize=0.78\hsize\epsffile{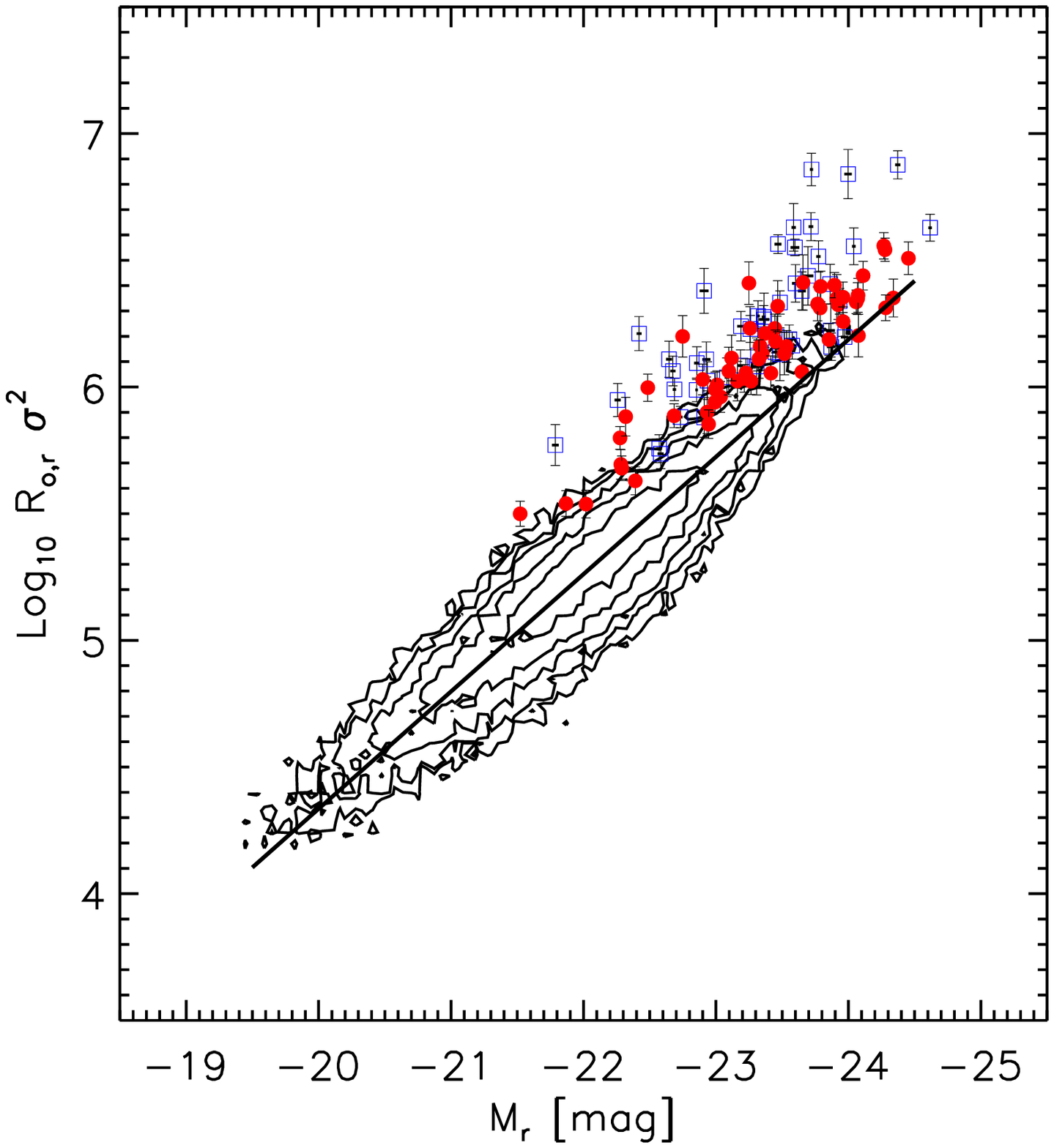}
 \epsfxsize=0.78\hsize\epsffile{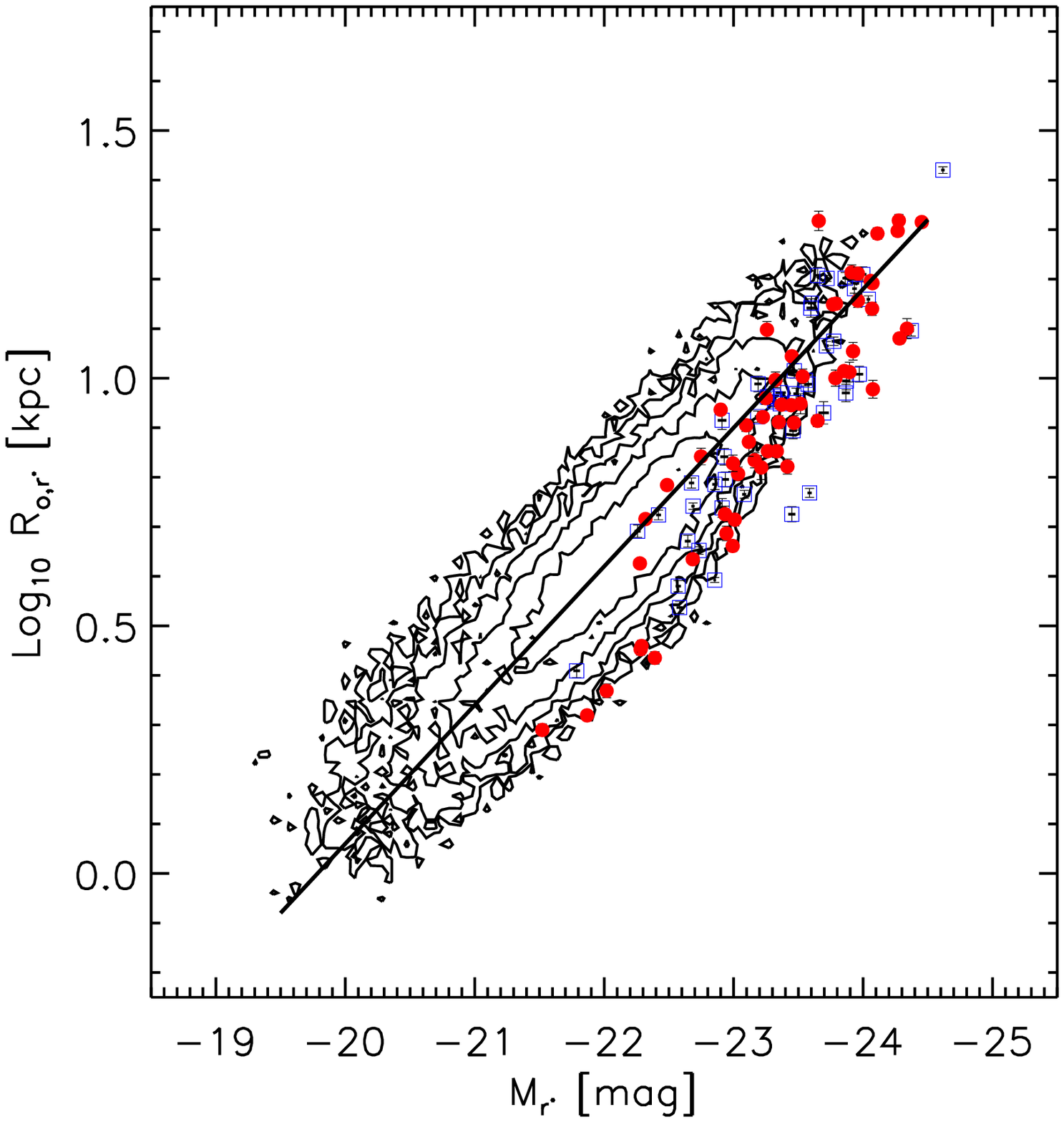}
 \epsfxsize=0.78\hsize\epsffile{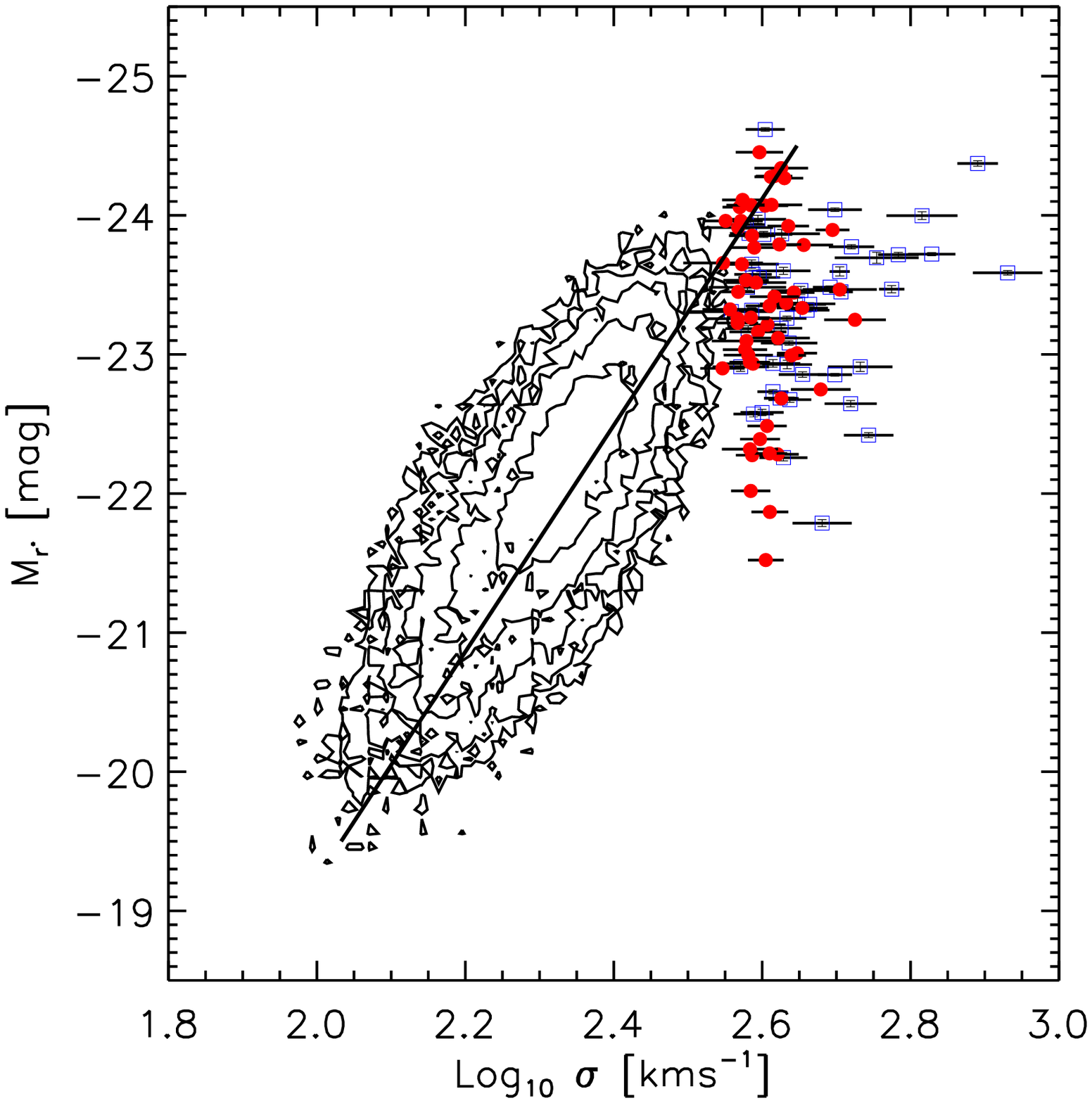}
\vspace{3.5cm}
 \label{scalings}
\end{figure}

The objects with $\sigma>350$~km~s$^{-1}$ which are not obvious doubles 
(red circles) clearly are extremes:  they outline the high-mass border 
of the mass versus luminosity relation (top right panel), and they have 
the smallest sizes and the largest velocity dispersions for their 
luminosities (middle right and bottom right).  
However, although they outline the borders of these relations, 
they are not clear outliers.  They also outline the borders of 
the Fundamental Plane (top left) and the size--surface-brightness 
(middle left) relations, but they are not outliers.  
In addition, although they are amongst the reddest objects (bottom left), 
they are not redder than extrapolation of the color--$\sigma$ relation 
from smaller $\sigma$ would suggest.  In fact, these objects tend to 
lie slightly blueward of this relation, but again, they are not outliers.  
These objects appear to simply be the high velocity dispersion tail of 
the early-type galaxy population.  
The next section presents more evidence that many of these objects are 
not superpositions.  

In contrast, the objects classified as superpositions (blue squares) 
are clear outliers from some of the scaling relations.  In particular, 
they are offset towards extremely small sizes from the 
Fundamental Plane relation (top left).  They also tend to be clear 
outliers from the mass versus luminosity relation, being offset towards 
very large mass values (top right).  However, in the other scaling 
relations, they are not obviously different from the large-$\sigma$ 
objects which are not obvious superpositions, although they tend to 
scatter even more towards the bluer end of the color-$\sigma$ relation.  
In most cases, the offset from the scaling relations is removed if one 
assumes that because of the superposition, the correct position of the 
symbol is approximately given by making the luminosity fainter by a 
factor of two, decreasing the velocity dispersion by a factor of 
between 1.4 and 2 as well, but leaving the color unchanged.  (The 
factor of two in luminosity is easily justified, since if the 
luminosity ratio was extreme, the light from the fainter member of the 
pair would simply not be noticed.  The factor of 1.5 or so in $\sigma$ 
is less straightforward; it is approximately what we find in our 
analysis of the superpositions in Appendix~\ref{super}.)  

Perhaps the most striking feature of the different panels is that the 
objects which were not obvious doubles are also not obvious outliers 
from the scaling relations defined by the main early-type sample (although 
they do define the borders).  In constrast, the objects we classified as 
doubles are more distant outliers---even though these scaling relations 
played no role in the determining whether an object was a single or a 
double.  
Clearly, identifying outliers from the Fundamental Plane and mass versus 
luminosity relations is a simple way of searching for superpositions.

\begin{figure}[t]
 \centering
 \epsfxsize=\hsize\epsffile{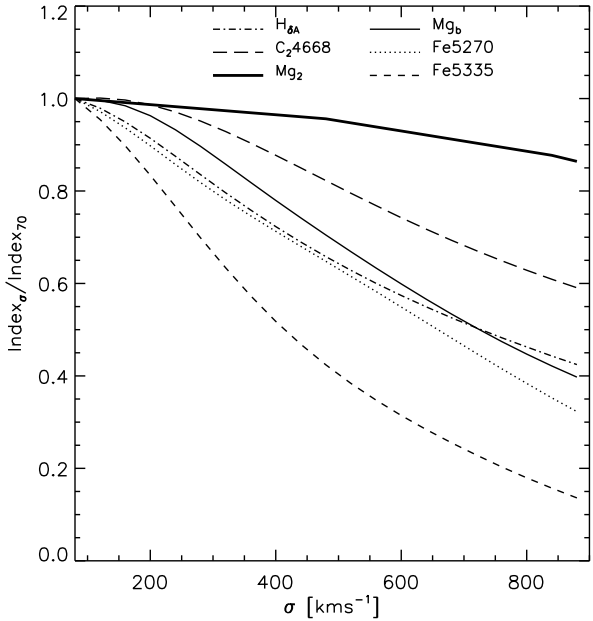}
 \caption{Velocity dispersion correction factor for various indices, 
          calibrated using the models of Bruzual \& Charlot (2003).  
          This correction does not depend on metallicity.}
 \label{IsigmaI80}
\end{figure}

\section{On rejecting the superposition hypothesis}\label{mg2test}
The previous section argued that doubles were relatively easy to identify, 
since they were obvious outliers from the scaling relations defined by 
the bulk of the population.  But is there a way to reject the 
superposition hypothesis?  In this section, we argue that this may 
be possible.  The idea is that if superposition tends to increase 
the estimated velocity dispersion, then it probably also changes 
the infered absorption line-strengths in the spectrum.  For instance, 
if two identical galaxies have a small line-of-sight separation, 
and all lines in the spectra of the individual objects have Gaussian 
profiles, then the spectrum of the superposition will have lines 
which are broader (hence increasing the infered $\sigma$).  Roughly 
speaking, line-strengths are related to the ratio of the flux in the 
line-center to the flux in the wings, so the superposition will have 
a smaller line-strength.  Therefore, superpositions may be obvious 
outliers from any linestrength$-\sigma$ relations defined by the bulk 
of the population.

\begin{figure}[t]
 \centering
 \epsfxsize=\hsize\epsffile{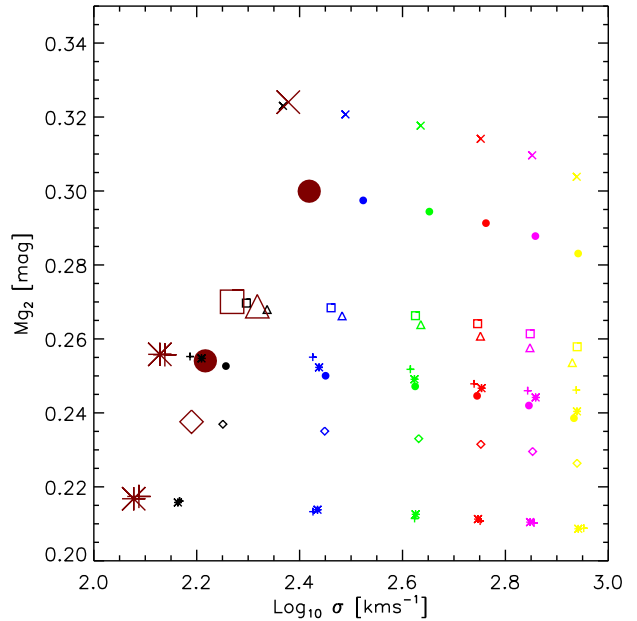}
 \caption{Effect of superposition on estimated velocity dispersions $\sigma$ 
          and Mg$_2$ line-strengths; the superposition is assumed to be 
          of otherwise identical objects.  Large (brown) symbols show the 
          values of $\sigma$ and Mg$_2$ for a single component (i.e., 
          line-of-sight separation $\Delta z=0$); small symbols show how 
          the estimated values change as the velocity separation between 
          the two components increases in steps of 200~km~s$^{-1}$:  
          $\sigma$ increases and Mg$_2$ decreases with increasing separation.  }
 \label{simMg2}
\end{figure}

With this in mind, we have measured various indicators of the 
chemical compositions of these objects:  
the Lick indices Mg$b$ and $\langle{\rm Fe}\rangle$ 
are sensitive to both age and metallicity, their ratio is an indicator 
of the relative abundances of $\alpha$-elements, 
$C_2 4668$ is also sensitive to age and metallicity, and 
H$\delta_{\rm A}$ and H$\gamma_{\rm F}$ are indicators of more recent 
star formation (Worthey \& Ottaviani 1997).  
It is conventional to report the measured value after  
correction for the effects of velocity dispersion and aperture: 
the correction factors for velocity dispersion are shown in 
Figure~\ref{IsigmaI80}.  Since $\sigma$ may be large, the corrections 
may also be large for some of the indices (e.g. the two iron lines 
Fe5270 and Fe5335).  However, the correction for Mg$_2$ is small; less 
than ten percent even when $\sigma\sim 600$~km~s$^{-1}$.  
For this reason, we will use it in what follows.  
In addition, the Mg$_2$ index can be measured accurately also for 
relatively low S/N spectra.

\begin{figure}[t]
 \centering
 \epsfxsize=0.74\hsize\epsffile{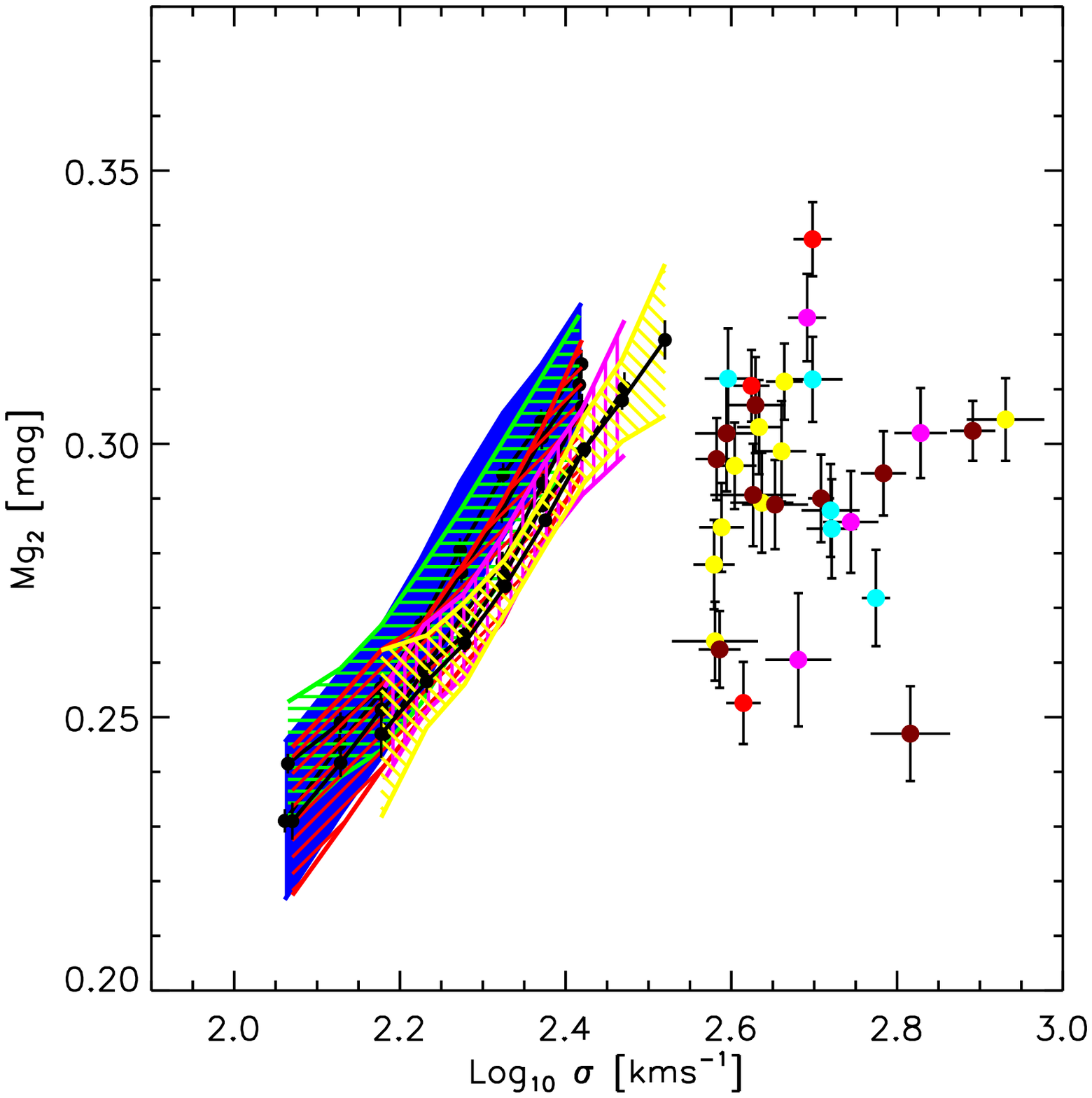}
 \epsfxsize=0.74\hsize\epsffile{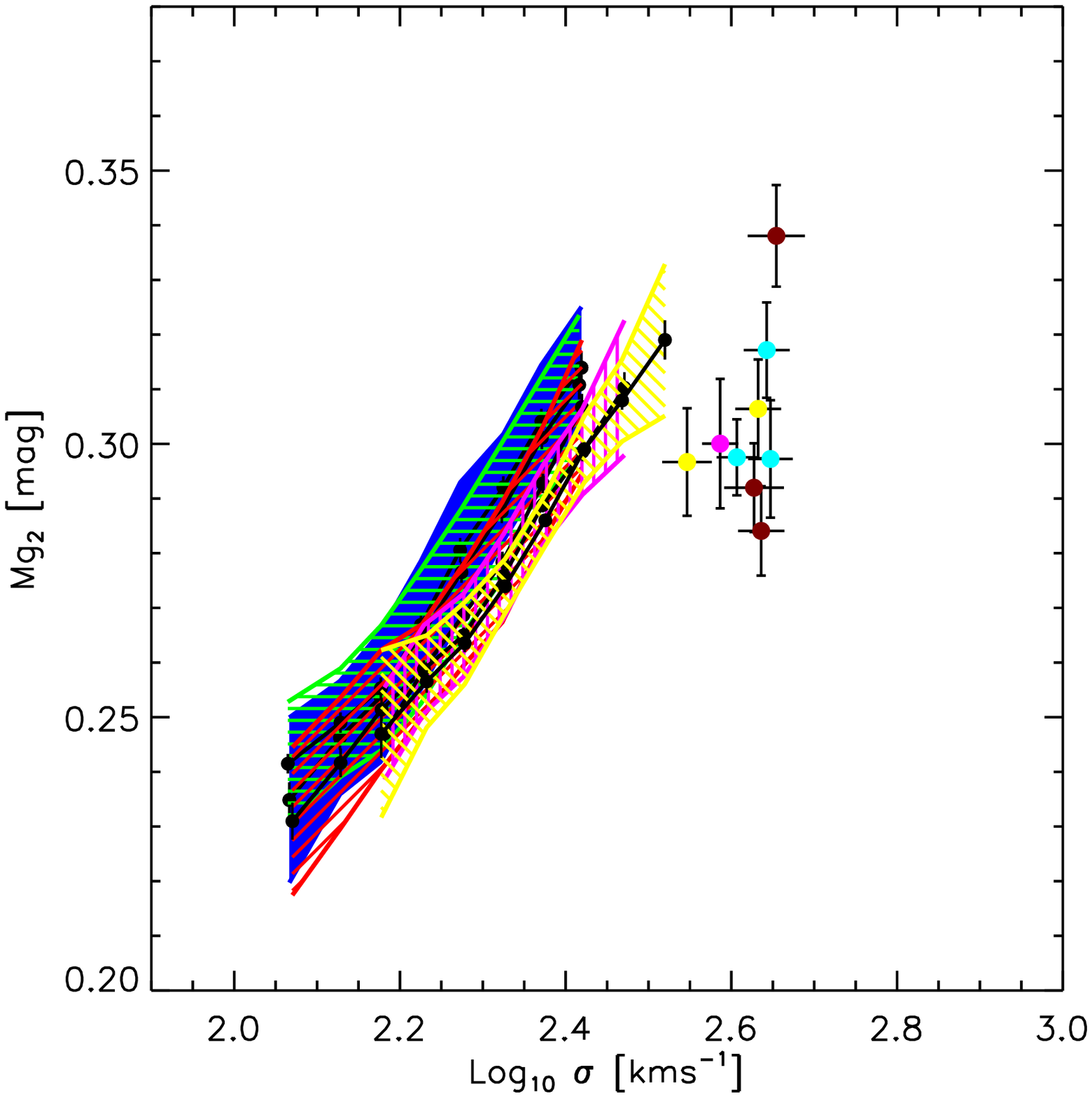}
 \epsfxsize=0.74\hsize\epsffile{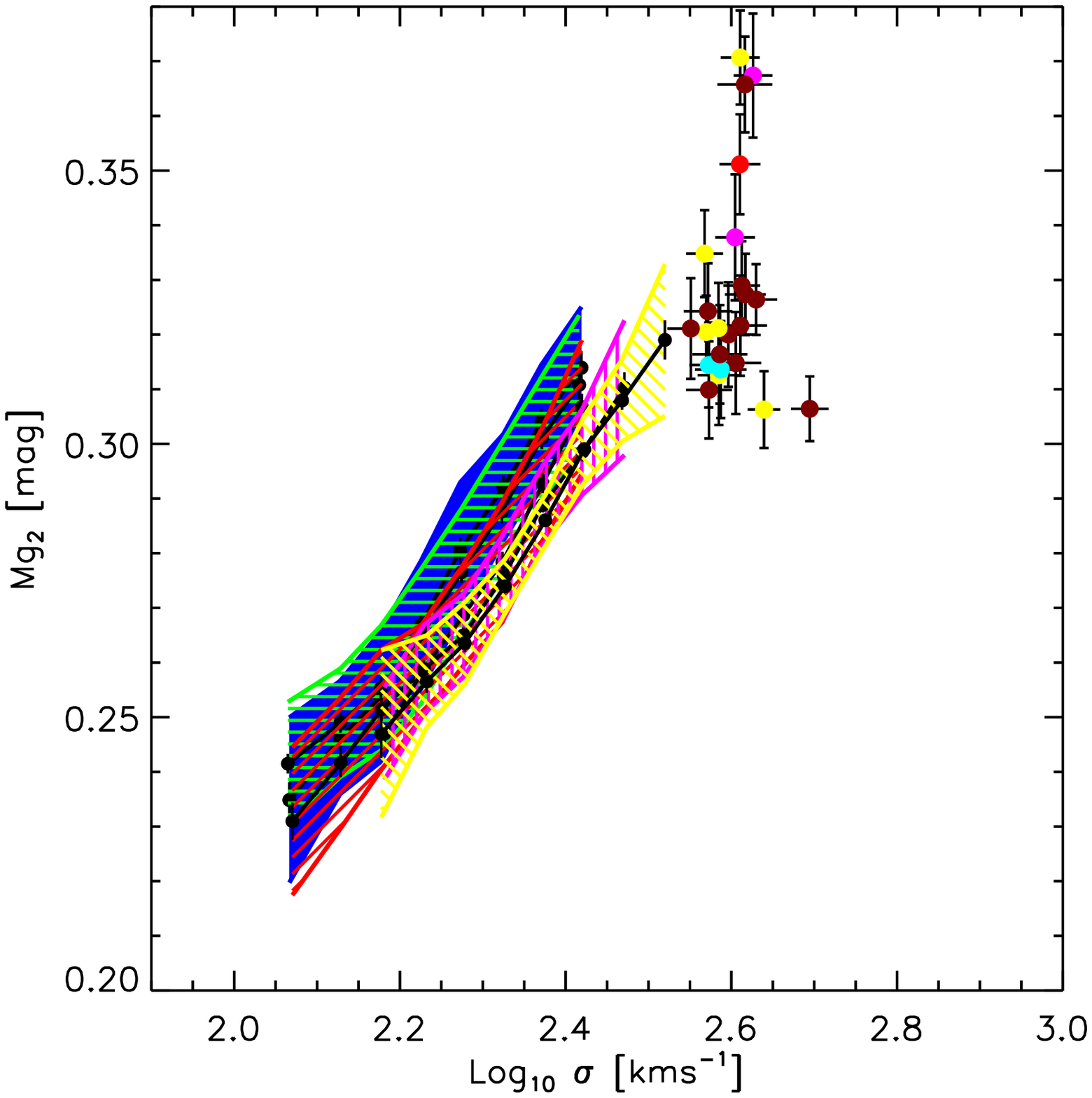}
 \caption{Comparison of Mg$_2-\sigma$ correlation defined by the bulk 
          of the population (hashed regions), with the locus of points 
          defined by the objects with strong, intermediate, and weak 
          evidence for superposition (top, middle and bottom panels).  
          Superpositions should be located down and to the right of the 
          true relation---this is true of most of the objects in the top 
          panel.  By this measure, objects in the bottom panel may well be 
          single massive galaxies. }
 \label{Mg2all}
\end{figure}

To study the effect of superposition on $\sigma$ and Mg$_2$, we have 
summed two identical spectra (chosen from among the high S/N composite 
spectra described in the next paragraph), with redshift separations 
$\Delta z$, and computed the velocity dispersion and Mg$_2$ strengths 
using the usual methods.  Figure~\ref{simMg2} shows the results of this 
exercise as $c\Delta z$ is varied from 0 to 1200~km~s$^{-1}$, in steps of 
200~km~s$^{-1}$.  Large (brown) symbols show $\sigma$ and line-strength 
when $\Delta z=0$ (so the values are the same as of the single component), 
and small symbols show the estimated $\sigma$ and Mg$_2$ value as the 
separation between the pair increases.  Increasing the separation tends 
to increase $\sigma$ and decrease Mg$_2$, as expected.  

Our use of identical spectra to illustrate this effect is less 
unrealistic than one might have thought.  
This is because the estimated velocity dispersion of the 
blend depends not only on the velocity dispersion and line-of-sight 
separation of the components, but also on the flux-ratio of the two 
components.  Hence, the two components should have similar fluxes, 
and, being at approximately the same redshift, they should have 
approximately the same luminosities.  However, velocity dispersion 
and luminosity are correlated:  in this sample, $L\propto \sigma^4$ 
(Bernardi et al. 2003b).  
If the individual components both lie on this $\sigma-L$ relation, 
then if the two values of $\sigma$ differ by more than a factor of 
$\sim 1.8$, the flux-ratio will be larger than $\sim 10$, 
and the smaller component is unlikely to have a significant effect.  
Thus, if we detect superpositions from the spectra, it is likely 
that the spectra of the two components will have relatively similar 
velocity dispersions, and hence relatively similar absorption features.  
This expectation is borne out by the fact that, of the systems which 
were clear superpositions, the estimated velocity dispersions of the 
two components tend to be similar (Figures~\ref{doubleimgspec}).  
Thus, even in the general case, the estimates of the effect of 
superposition on the Mg$_2-\sigma$ relation shown in 
Figure~\ref{simMg2} are likely to be qualitatively correct.  

To see if the Mg$_2-\sigma$ relation provides a reasonable diagnostic 
of superposition, we must compare the location of the systems with large 
$\sigma$ in the Mg$_2-\sigma$ plane, with the region populated by the 
main sample.  To define the Mg$_2-\sigma$ relation associated with the 
bulk of the early-type population, we used composite spectra constructed 
as described in Bernardi et al. (2005).  
Each composite has S/N $\sim 100$ or larger (making the estimate of the 
line-strength more reliable), and is made by summing the spectra of 
galaxies with similar magnitudes, sizes, velocity dispersions and redshifts.

The different panels of Figure~\ref{Mg2all} all show the Mg$_2-\sigma$ 
relation traced by the main population:  
black solid lines show the weighted mean of the index strength, 
hashed regions show the rms scatter around these mean values in the 
different redshift bins (blue, green, red, magenta and yellow bands 
show results for $0.04 < z < 0.07$, $0.07 < z < 0.09$, $0.09 < z < 0.12$, 
$0.12 < z < 0.15$, and $0.15 < z < 0.20$; colors of the filled circles in 
the different panels also indicate these same redshift bins. Cyan and brown 
symbols represent galaxies at even higher redshifts:  $0.2 \le z < 0.25$ and 
$z \ge 0.25$, respectively).  
Notice that the Mg$_2-\sigma$ relation evolves:  at fixed $\sigma$, 
Mg$_2$ decreases with increasing $z$.  This is consistent with previous 
work (e.g. Bernardi et al. 2003d).  
In the top panel, the filled circles with error bars represent those 
objects classified as superpositions with S/N $\ge 18$ 
(it is difficult to make reliable estimates at smaller S/N ratios).  
The previous figure suggests that superpositions should lie down and 
to the right of the true relation:  the filled circles do indeed show 
such an offset.  
The filled circles in the middle panel show those objects which are not 
obvious superpositions, but do still have odd features in their spectra.  
While the objects which lie below the Mg$_2-\sigma$ relation are 
probably superpositions, the ones which lie close to the relation may 
well be single objects.  
The bottom panel shows a similar analysis applied to the objects for which 
the evidence for superposition is weakest:  neither the images nor the 
spectra showed compelling evidence for superposition.  
These objects tend to lie on or slightly above the mean Mg$_2-\sigma$ 
relation; the results of Figure~\ref{simMg2} suggest that they are 
very unlikely to be superpositions.  These objects may well be some 
of the most massive galaxies in the Universe.  

\section{Discussion}\label{discuss}
We searched the SDSS database for a population of objects with 
anomalously large velocity dispersions, and found $\sim 100$ objects 
with estimated dispersions in excess of 350~km~s$^{-1}$.  
Of these, about half appear to be superpositions (so the reported 
velocity dispersion is unrealistic); in many cases, the evidence for 
superposition comes not from the images but from the spectra 
(Figure~\ref{doubleimgspec}).  
These superpositions are rare: analytic and Monte-Carlo analyses in 
Appendix~\ref{super} suggest that one in every three hundred objects 
should have a neighbor within one arcsec.  Moreover, of alignments 
closer than one arcsec, not more than ten percent are expected to be 
from objects in different groups.  If the superpositions we see really 
are in the same halos, then our estimates of the line-of-sight 
separations imply halo masses of order $5\times 10^{14}h^{-1}M_\odot$.  

The large-$\sigma$ objects which are not obviously superpositions populate 
the tails of the scaling relations defined by the bulk of the early-type 
galaxy population (quantified by Bernardi et al. 2005), but they are 
not distant outliers from these relations (Figures~\ref{scalings} 
and~\ref{Mg2all}).  
Moreover, if only half of these objects turn out to be superpositions, 
and the other half are indeed single galaxies, then the abundance of 
singles is not inconsistent with the number expected by extrapolation 
of the observed abundance of smaller systems (from Sheth et al. 2003).  

If these large-$\sigma$ objects are indeed massive galaxies, and the 
velocity dispersions do reflect virial equilibrium motions, then it 
might be worth searching for evidence of gravitational lensing around 
these objects: they would have Einstein radii  
$4\pi (\sigma/c)^2 (d_{\rm ls}/d_{\rm os})\sim 2.3\arcsec 
(d_{\rm ls}/d_{\rm os})/(1/2)$ $(\sigma/400~{\rm km~s}^{-1})^2$.
If they host black holes whose masses fall on the same mass-velocity 
dispersion relation as is seen locally, 
 $M_{\rm BH}/10^9M_\odot = 2\,(\sigma/400~{\rm km~s}^{-1})^4$ 
(Gebhardt et al. 2000; Ferrarese \& Merritt 2000; Tremaine et al. 2002), 
then the black-holes are enormous indeed.  
In this case, it will be interesting to see if the light-profile shows 
any evidence of the black-hole in the center (e.g. Lauer et al. 2002) 
using HST.  Because they are large and luminous, these objects should 
be relatively easy targets.  Therefore, it should also be possible to 
measure spatially resolved velocity dispersions from ground-based 
facilities.  

The superpositions are interesting in their own right.  
The abundance of strong gravitational lenses has been used to place 
limits on the geometry of the Universe (e.g. Mitchell et al. 2004).  
However, the observed distributions of image multiplicities, separations 
and flux ratios are difficult to reconcile with single-component lens 
models.  This has led to some interest in the properties of lenses with 
multiple components (e.g., Rusin \& Tegmark 2001; Cohn \& Kochanek 2004).  
Since early-type galaxies are expected to be the dominant lens population, 
the distribution of pair separations and velocity dispersions in our 
catalog of superpositions can be used to incorporate realistic lens pairs 
into models of binary-lenses.  This is the subject of work in progress.  

And finally, in principle, the number and spatial distribution of close 
superpositions contains information about the time-scale of mergers.  
In this regard, it is interesting that a number of the objects in our 
sample appear to have slightly peculiar morphologies.  If this reflects 
a recent merger, then it is interesting to recall that none of the 
spectra in our sample show strong emission lines.  Therefore, it may 
be that these objects are the low redshift analogs of the red 
interacting galaxies seen in the GOODS survey (Somerville et al. 2003).  
Or perhaps they are fossil groups of the sort discussed by 
Vikhlinin (1999) and Jones et al. (2003).  
Follow-up observations of these objects is ongoing.

\vspace{0.5cm}

We thank Karl Gebhardt for encouragement.

Funding for the creation and distribution of the SDSS Archive has
been provided by the Alfred P. Sloan Foundation, the Participating
Institutions, the National Aeronautics and Space Administration, the
National Science Foundation, the U.S. Department of Energy, the
Japanese Monbukagakusho, and the Max Planck Society. The SDSS Web site
is http://www.sdss.org/. 

The SDSS is managed by the Astrophysical Research Consortium (ARC)
for the Participating Institutions. The Participating Institutions are
The University of Chicago, Fermilab, the Institute for Advanced Study,
the Japan Participation Group, The Johns Hopkins University, the Korean
Scientist Group, Los Alamos National Laboratory, the
Max-Planck-Institute for Astronomy (MPIA), the Max-Planck-Institute
for Astrophysics (MPA), New Mexico State University, University of
Pittsburgh, Princeton University, the United States Naval Observatory,
and the University of Washington.

\begin{table*}
\caption[]{Physical parameters and median errors of the objects in our 
           sample which are not obvious superpositions.  Estimates of 
           Mg$_2$ were only made if S/N $\ge 18$.  Asterisk in final 
           column denotes objects for which the evidence for superposition 
           is weakest.  \\}
\small
\begin{tabular}{ccccccccccccl}
 \hline &&&\\
  name & $z$ & $M_r$ & $e_M$ & 
  $g-r$ & $e_{g-r}$ & log$_{10} R$ & $e_R$ &
  $\sigma$ & $e_{\sigma}$ & Mg$_2$ & $e_{\rm{Mg}_2}$ & $S/N$ \\
  & & [mag] & [mag] & [mag] & [mag] & [kpc] & [kpc] & kms$^{-1}$ & kms$^{-1}$ & [mag] & [mag] & \\
\hline &&&\\
SDSS J094035.8+022950.0 & 0.15196 & $-$22.90 & 0.02 & 0.80 & 0.02 &  0.93 &  0.01 & 352 &  29 & 0.297 & 0.010 &  18 \\
SDSS J112626.6+003620.7 & 0.28655 & $-$23.66 & 0.04 & 0.85 & 0.06 &  1.32 &  0.02 & 352 &  53 &  --   &  --   &  12 \\
SDSS J083551.2+392621.7 & 0.26035 & $-$23.96 & 0.02 & 0.84 & 0.02 &  1.13 &  0.01 & 355 &  29 & 0.321 & 0.009 &  19*\\
SDSS J013431.5+131436.4 & 0.23949 & $-$23.32 & 0.03 & 0.81 & 0.04 &  1.00 &  0.02 & 360 &  37 &  --   &  --   &  13*\\
SDSS J162332.4+450032.0 & 0.19827 & $-$23.23 & 0.02 & 0.88 & 0.02 &  0.90 &  0.01 & 368 &  28 &  --   &  --   &  16*\\
SDSS J132808.5+031817.1 & 0.22034 & $-$23.26 & 0.03 & 0.74 & 0.03 &  1.09 &  0.02 & 369 &  41 &  --   &  --   &  13 \\
SDSS J010803.2+151333.6 & 0.16773 & $-$23.45 & 0.01 & 0.86 & 0.02 &  1.04 &  0.01 & 369 &  22 & 0.335 & 0.008 &  21*\\
SDSS J083445.2+355142.0 & 0.32787 & $-$23.91 & 0.03 & 0.92 & 0.05 &  1.18 &  0.02 & 371 &  52 &  --   &  --   &  13 \\
SDSS J131419.7$-$012726.0 & 0.18011 & $-$24.06 & 0.01 & 0.85 & 0.01 &  1.20 &  0.01 & 371 &  23 & 0.320 & 0.007 &  24* \\
SDSS J091944.2+562201.1 & 0.27775 & $-$23.96 & 0.02 & 0.94 & 0.03 &  1.20 &  0.01 & 373 &  30 & 0.324 & 0.009 &  19*\\
SDSS J155944.2+005236.8 & 0.21356 & $-$23.65 & 0.02 & 0.87 & 0.03 &  0.90 &  0.01 & 373 &  22 & 0.314 & 0.008 &  22*\\
SDSS J135602.4+021044.6 & 0.26271 & $-$24.11 & 0.02 & 0.91 & 0.04 &  1.30 &  0.01 & 374 &  29 & 0.310 & 0.009 &  19*\\
SDSS J075923.1+274148.3 & 0.19477 & $-$23.03 & 0.02 & 0.95 & 0.01 &  0.82 &  0.01 & 376 &  30 &  --   &  --   &  17*\\
SDSS J141341.4+033104.3 & 0.24639 & $-$23.53 & 0.03 & 0.88 & 0.04 &  0.99 &  0.02 & 379 &  33 &  --   &  --   &  16*\\
SDSS J112842.0+043221.7 & 0.20533 & $-$22.99 & 0.03 & 0.93 & 0.04 &  0.83 &  0.02 & 381 &  34 &  --   &  --   &  14*\\
SDSS J124134.3+604147.2 & 0.23336 & $-$23.10 & 0.02 & 0.78 & 0.03 &  0.84 &  0.01 & 381 &  47 &  --   &  --   &  14 \\
SDSS J093124.4+574926.6 & 0.22792 & $-$22.94 & 0.03 & 0.92 & 0.04 &  0.69 &  0.02 & 383 &  27 &  --   &  --   &  16*\\
SDSS J103344.2+043143.5 & 0.15939 & $-$22.32 & 0.02 & 0.95 & 0.02 &  0.72 &  0.01 & 383 &  41 &  --   &  --   &  11 \\
SDSS J221414.3+131703.7 & 0.15335 & $-$22.02 & 0.03 & 0.82 & 0.01 &  0.37 &  0.02 & 384 &  28 & 0.321 & 0.008 &  20*\\
SDSS J225331.3+130116.9 & 0.19812 & $-$23.26 & 0.02 & 0.83 & 0.02 &  0.85 &  0.01 & 384 &  26 & 0.313 & 0.009 &  19*\\
SDSS J120011.1+680924.8 & 0.26275 & $-$24.07 & 0.02 & 0.88 & 0.03 &  1.18 &  0.01 & 385 &  34 & 0.316 & 0.009 &  19*\\
SDSS J154651.5+570736.2 & 0.12845 & $-$22.28 & 0.02 & 0.85 & 0.01 &  0.62 &  0.01 & 385 &  24 & 0.300 & 0.012 &  22 \\
SDSS J211019.2+095047.1 & 0.23073 & $-$23.85 & 0.02 & 0.90 & 0.03 &  0.99 &  0.01 & 386 &  32 & 0.314 & 0.009 &  19*\\
SDSS J160239.1+022110.0 & 0.21930 & $-$22.93 & 0.03 & 0.88 & 0.03 &  0.73 &  0.02 & 386 &  41 &  --   &  --   &  10 \\
SDSS J154017.3+430024.5 & 0.25370 & $-$23.77 & 0.02 & 0.97 & 0.03 &  1.14 &  0.01 & 388 &  36 &  --   &  --   &  16*\\
SDSS J111525.7+024033.9 & 0.28489 & $-$23.52 & 0.04 & 0.88 & 0.04 &  0.91 &  0.02 & 391 &  44 &  --   &  --   &  13 \\
SDSS J130615.8+600125.2 & 0.26580 & $-$23.16 & 0.03 & 0.93 & 0.05 &  0.84 &  0.02 & 392 &  44 &  --   &  --   &  12 \\
SDSS J145506.8+615809.7 & 0.27422 & $-$24.45 & 0.02 & 0.87 & 0.02 &  1.32 &  0.01 & 394 &  36 & 0.320 & 0.010 &  18*\\
SDSS J235354.1$-$093908.3 & 0.18764 & $-$22.39 & 0.02 & 0.82 & 0.04 &  0.44 &  0.02 & 395 &  27 &  --   &  --   &  16*\\
SDSS J082216.5+481519.1 & 0.12705 & $-$21.52 & 0.02 & 0.80 & 0.02 &  0.29 &  0.01 & 402 &  28 & 0.338 & 0.012 &  19*\\
SDSS J124609.4+515021.6 & 0.26965 & $-$24.07 & 0.02 & 0.85 & 0.02 &  1.12 &  0.01 & 402 &  35 & 0.315 & 0.009 &  18* \\
SDSS J204712.0$-$054336.7 & 0.14386 & $-$22.49 & 0.02 & 0.80 & 0.01 &  0.78 &  0.01 & 404 &  32 &  --   &  --   &  17 \\
SDSS J085738.8+561121.0 & 0.24427 & $-$23.21 & 0.04 & 0.81 & 0.04 &  0.82 &  0.03 & 404 &  25 & 0.298 & 0.007 &  21 \\
SDSS J151741.7$-$004217.6 & 0.11610 & $-$21.87 & 0.02 & 0.82 & 0.01 &  0.32 &  0.01 & 407 &  27 & 0.351 & 0.009 &  24*\\
SDSS J082646.7+495211.5 & 0.16037 & $-$22.29 & 0.02 & 0.91 & 0.02 &  0.46 &  0.01 & 408 &  26 & 0.371 & 0.009 &  21*\\
SDSS J011613.8$-$092625.2 & 0.26262 & $-$23.35 & 0.02 & 0.92 & 0.06 &  0.88 &  0.02 & 408 &  39 & 0.322 & 0.009 &  18*\\
SDSS J084257.5+362159.3 & 0.28227 & $-$24.28 & 0.02 & 0.98 & 0.03 &  1.28 &  0.01 & 409 &  31 & 0.329 & 0.008 &  21*\\
SDSS J204642.1+000507.7 & 0.25658 & $-$23.42 & 0.03 & 0.97 & 0.06 &  0.82 &  0.02 & 413 &  35 & 0.366 & 0.009 &  19*\\
SDSS J171328.4+274336.6 & 0.29718 & $-$24.28 & 0.02 & 0.94 & 0.04 &  1.08 &  0.01 & 413 &  27 & 0.327 & 0.007 &  22*\\
SDSS J134126.7+013641.1 & 0.38403 & $-$24.08 & 0.03 & 0.81 & 0.07 &  0.86 &  0.02 & 414 &  49 &  --   &  --   &  16 \\
SDSS J224248.8+135430.8 & 0.18549 & $-$22.28 & 0.02 & 0.77 & 0.02 &  0.47 &  0.01 & 417 &  30 &  --   &  --   &  17 \\
SDSS J114747.0+034838.7 & 0.26737 & $-$23.12 & 0.05 & 0.90 & 0.07 &  0.82 &  0.03 & 419 &  62 &  --   &  --   &  12 \\
SDSS J135533.4+515617.8 & 0.27669 & $-$23.79 & 0.02 & 0.87 & 0.04 &  1.13 &  0.01 & 420 &  43 &  --   &  --   &  14 \\
SDSS J133724.7+033656.5 & 0.13343 & $-$22.68 & 0.01 & 0.83 & 0.01 &  0.63 &  0.01 & 422 &  31 & 0.367 & 0.011 &  19*\\
SDSS J031539.2$-$081014.3 & 0.34977 & $-$24.34 & 0.04 & 0.82 & 0.09 &  1.06 &  0.02 & 423 &  38 & 0.292 & 0.008 &  22 \\
SDSS J104056.4$-$010358.7 & 0.25026 & $-$24.27 & 0.02 & 0.79 & 0.02 &  1.30 &  0.01 & 426 &  30 & 0.326 & 0.006 &  25*\\
SDSS J141922.4+011457.8 & 0.16984 & $-$23.37 & 0.02 & 0.81 & 0.02 &  0.95 &  0.01 & 428 &  30 & 0.306 & 0.009 &  19 \\
SDSS J133046.1+585049.9 & 0.31075 & $-$23.92 & 0.03 & 0.85 & 0.03 &  1.03 &  0.02 & 432 &  38 & 0.284 & 0.008 &  20 \\
SDSS J161541.3+471004.3 & 0.19766 & $-$22.99 & 0.02 & 0.81 & 0.00 &  0.66 &  0.01 & 435 &  21 & 0.306 & 0.007 &  24*\\
SDSS J132356.8+001049.8 & 0.22705 & $-$23.44 & 0.02 & 0.86 & 0.02 &  0.95 &  0.01 & 439 &  32 & 0.317 & 0.009 &  20 \\
SDSS J111505.5+051833.6 & 0.21901 & $-$23.01 & 0.02 & 0.79 & 0.04 &  0.71 &  0.01 & 443 &  31 & 0.297 & 0.011 &  18 \\
SDSS J161615.5+435559.5 & 0.25178 & $-$23.33 & 0.02 & 0.81 & 0.04 &  0.85 &  0.01 & 451 &  40 & 0.338 & 0.009 &  19 \\
SDSS J232331.4$-$102551.7 & 0.29211 & $-$23.79 & 0.03 & 0.91 & 0.08 &  0.99 &  0.02 & 453 &  46 &  --   &  --   &  17 \\
SDSS J141102.6+030805.7 & 0.18917 & $-$22.75 & 0.03 & 0.78 & 0.02 &  0.84 &  0.02 & 477 &  48 &  --   &  --   &  12 \\
SDSS J032834.7+001050.1 & 0.31557 & $-$23.89 & 0.04 & 0.93 & 0.07 &  1.02 &  0.02 & 494 &  28 & 0.306 & 0.006 &  30*\\
SDSS J223859.6+004041.4 & 0.27472 & $-$23.47 & 0.03 & 0.83 & 0.04 &  0.86 &  0.02 & 507 &  66 &  --   &  --   &  13 \\
SDSS J010354.1+144814.1 & 0.22693 & $-$23.25 & 0.02 & 0.96 & 0.06 &  0.97 &  0.01 & 529 &  58 &  --   &  --   &  16 \\
\hline &&&\\
\end{tabular}
\label{tab:singles} 
\end{table*}
\normalsize

\begin{table*}
\caption[]{Physical parameters of the objects in our sample which are almost 
           certainly superpositions.  An estimate of the separation along 
           the line-of-sight is also provided. For these objects, the 
measured parameters of the blend are almost certainly not those of the 
individual components, so the measurement errors are not reported.\\}
\small
\begin{tabular}{ccccccccc}
 \hline &&&\\
  name & $z$ & $M_r$ &  
  $g-r$ & log$_{10} R$ &
  $\Delta cz$ & $\sigma$ & Mg$_2$ & $S/N$\\
  & & [mag] & [mag] & [kpc] & [km s$^{-1}$] & [km s$^{-1}$] & [mag] &  \\
\hline &&&\\
SDSS J150821.4$-$021637.9 & 0.27380 & $-$23.31 & 0.90 &  0.97 &   900 &  361 &  --   &  12\\
SDSS J021046.9+143448.4 & 0.20529 & $-$22.91 & 0.82 &  0.74 &   600 &  371 &  --   &  15\\
SDSS J150128.7+033630.4 & 0.18459 & $-$23.93 & 0.79 &  1.18 &     -- &  379 & 0.278 &  20\\
SDSS J091545.5+505424.4 & 0.18364 & $-$23.48 & 0.81 &  0.97 &   900 &  380 & 0.264 &  21\\
SDSS J163005.3+463611.2 & 0.22330 & $-$23.19 & 0.78 &  0.90 &   750 &  380 &  --   &  16\\
SDSS J095937.6+031001.7 & 0.30452 & $-$23.87 & 0.86 &  0.99 &   750 &  382 & 0.297 &  22\\
SDSS J022433.2$-$073111.0 & 0.27716 & $-$23.58 & 0.93 &  1.03 &   600 &  385 & 0.262 &  23\\
SDSS J145710.1+604207.2 & 0.22112 & $-$23.32 & 0.87 &  0.95 &  1200 &  385 &  --   &  16\\
SDSS J144525.1+591327.5 & 0.29959 & $-$23.65 & 0.91 &  1.16 &    -- &  386 &  --   &  15\\
SDSS J140836.5+613108.0 & 0.16440 & $-$22.57 & 0.76 &  0.58 &   600 &  387 & 0.285 &  20\\
SDSS J142437.2+000835.6 & 0.32261 & $-$23.97 & 0.89 &  1.01 &    -- &  392 & 0.302 &  18\\
SDSS J135331.2+533430.7 & 0.22495 & $-$23.55 & 0.76 &  1.00 &   450 &  394 & 0.312 &  20\\
SDSS J173820.0+551638.8 & 0.20645 & $-$22.58 & 0.85 &  0.51 &   450 &  398 &  --   &  17\\
SDSS J152333.3+450335.7 & 0.25304 & $-$23.86 & 0.84 &  1.20 &    -- &  400 &  --   &  14\\
SDSS J215541.9+123128.6 & 0.19300 & $-$24.62 & 0.81 &  1.43 &   600 &  401 & 0.296 &  21\\
SDSS J120439.0+601211.2 & 0.11842 & $-$22.73 & 0.74 &  0.66 &   750 &  411 & 0.253 &  30\\
SDSS J231543.5$-$000511.6 & 0.22490 & $-$22.94 & 0.70 &  0.79 &   299 &  411 &  --   &  15\\
SDSS J153603.4+003749.3 & 0.09460 & $-$22.69 & 0.77 &  0.74 &   600 &  420 & 0.311 &  29\\
SDSS J104940.3+050307.1 & 0.31160 & $-$23.87 & 0.90 &  0.97 &   750 &  422 & 0.291 &  19\\
SDSS J122051.1+533436.2 & 0.20808 & $-$23.19 & 0.76 &  0.97 &   600 &  423 &  --   &  16\\
SDSS J020556.7+000056.6 & 0.17308 & $-$22.26 & 0.81 &  0.73 &   600 &  423 &  --   &  15\\
SDSS J142543.5+620500.0 & 0.25925 & $-$23.60 & 0.90 &  1.14 &   450 &  425 & 0.307 &  20\\
SDSS J141557.6+031821.2 & 0.16520 & $-$23.26 & 0.79 &  0.96 &   600 &  429 & 0.303 &  19\\
SDSS J075527.6+360749.6 & 0.24205 & $-$22.93 & 0.83 &  0.82 &   750 &  430 &  --   &  16\\
SDSS J102618.9+492119.2 & 0.19533 & $-$23.08 & 0.83 &  0.75 &   600 &  433 & 0.289 &  18\\
SDSS J214141.6+011146.9 & 0.16447 & $-$22.67 & 0.81 &  0.78 &   750 &  433 &  --   &  17\\
SDSS J074224.6+305345.6 & 0.29015 & $-$23.36 & 0.86 &  0.97 &    -- &  444 &  --   &  12\\
SDSS J021148.2+001639.7 & 0.29476 & $-$23.46 & 0.61 &  0.88 &   450 &  449 & 0.289 &  19\\
SDSS J114634.4+022147.5 & 0.19317 & $-$22.86 & 0.79 &  0.77 &   600 &  451 &  --   &  17\\
SDSS J014157.5$-$010626.3 & 0.15613 & $-$23.32 & 0.78 &  0.96 &   750 &  457 & 0.299 &  20\\
SDSS J133153.5+031750.5 & 0.17934 & $-$23.36 & 0.76 &  0.95 &   750 &  461 & 0.311 &  22\\
SDSS J153228.9+023916.5 & 0.13003 & $-$21.79 & 0.79 &  0.41 &   570 &  479 & 0.261 &  18\\
SDSS J104907.2+551314.9 & 0.12631 & $-$23.48 & 0.82 &  0.95 &   900 &  491 & 0.323 &  27\\
SDSS J000740.4+144506.7 & 0.11610 & $-$22.85 & 0.84 &  0.59 &   750 &  498 & 0.337 &  32\\
SDSS J152242.7+574009.5 & 0.20218 & $-$24.04 & 0.88 &  1.16 &  1050 &  498 & 0.312 &  20\\
SDSS J115514.0$-$012041.1 & 0.27735 & $-$23.60 & 0.84 &  1.15 &   750 &  506 &  --   &  12\\
SDSS J155614.1+484706.1 & 0.25938 & $-$23.45 & 0.79 &  0.71 &  1050 &  510 & 0.290 &  22\\
SDSS J090321.9+520607.6 & 0.21608 & $-$22.65 & 0.90 &  0.66 &  1050 &  524 & 0.288 &  21\\
SDSS J150550.3+042909.3 & 0.21763 & $-$23.77 & 0.80 &  1.05 &  1200 &  526 & 0.284 &  20\\
SDSS J130543.9+674615.6 & 0.21956 & $-$22.91 & 0.76 &  0.92 &   900 &  541 &  --   &  13\\
SDSS J131616.2+051403.7 & 0.14727 & $-$22.42 & 0.84 &  0.72 &   --  &  554 & 0.286 &  19\\
SDSS J094436.5+614411.5 & 0.33489 & $-$23.69 & 0.93 &  0.96 &   900 &  566 &  --   &  13\\
SDSS J080413.1+372737.9 & 0.23550 & $-$23.47 & 0.81 &  1.01 &   900 &  594 & 0.272 &  18\\
SDSS J162255.0+455514.7 & 0.25767 & $-$23.72 & 0.80 &  1.07 &  1050 &  607 & 0.295 &  22\\
SDSS J133035.4+590117.3 & 0.31102 & $-$24.00 & 0.88 &  1.18 &  1350 &  654 & 0.247 &  19\\
SDSS J103836.6+011749.4 & 0.12871 & $-$23.72 & 0.79 &  1.20 &  1200 &  673 & 0.302 &  27\\
SDSS J080234.9+362100.9 & 0.29396 & $-$24.37 & 0.87 &  1.08 &  1350 &  778 & 0.302 &  30\\
SDSS J021515.5+005823.7 & 0.19985 & $-$23.59 & 0.81 &  0.77 &  1350 &  852 & 0.304 &  23\\
\hline &&&\\
\end{tabular}
\label{tab:doubles} 
\end{table*}
{}

\appendix
\section{A. Superpositions}\label{super}

\subsection{A.1 Likelihood of superpositions}\label{like}
In the main text we stated that a number of our objects were 
superpositions, even though the SDSS photometric pipeline classified 
them as single objects.  In this section, we make a number of estimates 
of the likelihood of having a superposition with image separations of 
order 1 arcsec:  this is a convenient number as it happens to be 
approximately the size of the typical SDSS seeing disk, and it is 
slightly smaller than the 1.5 arcsecond radius of an SDSS fiber.  

A simple but inaccurate estimate follows from ignoring the fact that 
galaxies cluster.  If the galaxies were Poisson-distributed, then the 
typical number of galaxies within a small angle $\theta$ would be 
$\bar N = \bar n \pi\theta^2$, where $\bar n$ is the mean density of 
galaxies on the surface of the sky.  
For a Poisson distribution, the probability that a region of size $A$ 
which is centred on a randomly chosen galaxy contains at least one other 
galaxy is $p(>0|A) = 1 - {\rm exp}(-\bar N)$, where $\bar N$ denotes the 
typical number of galaxies in regions of size $A$.  
We are interested in regions which are typically an arcsecond in radius 
(recall that the diameter of an SDSS fiber is 3 arcsec); we will argue 
below that this means $\bar N\ll 1$, so $p(>0|A) \approx \bar N$.  
To justify this, we must estimate $\bar n$.  

\setcounter{figure}{0}

\begin{figure*}
 \centering
 \epsfxsize=0.9\hsize\epsffile{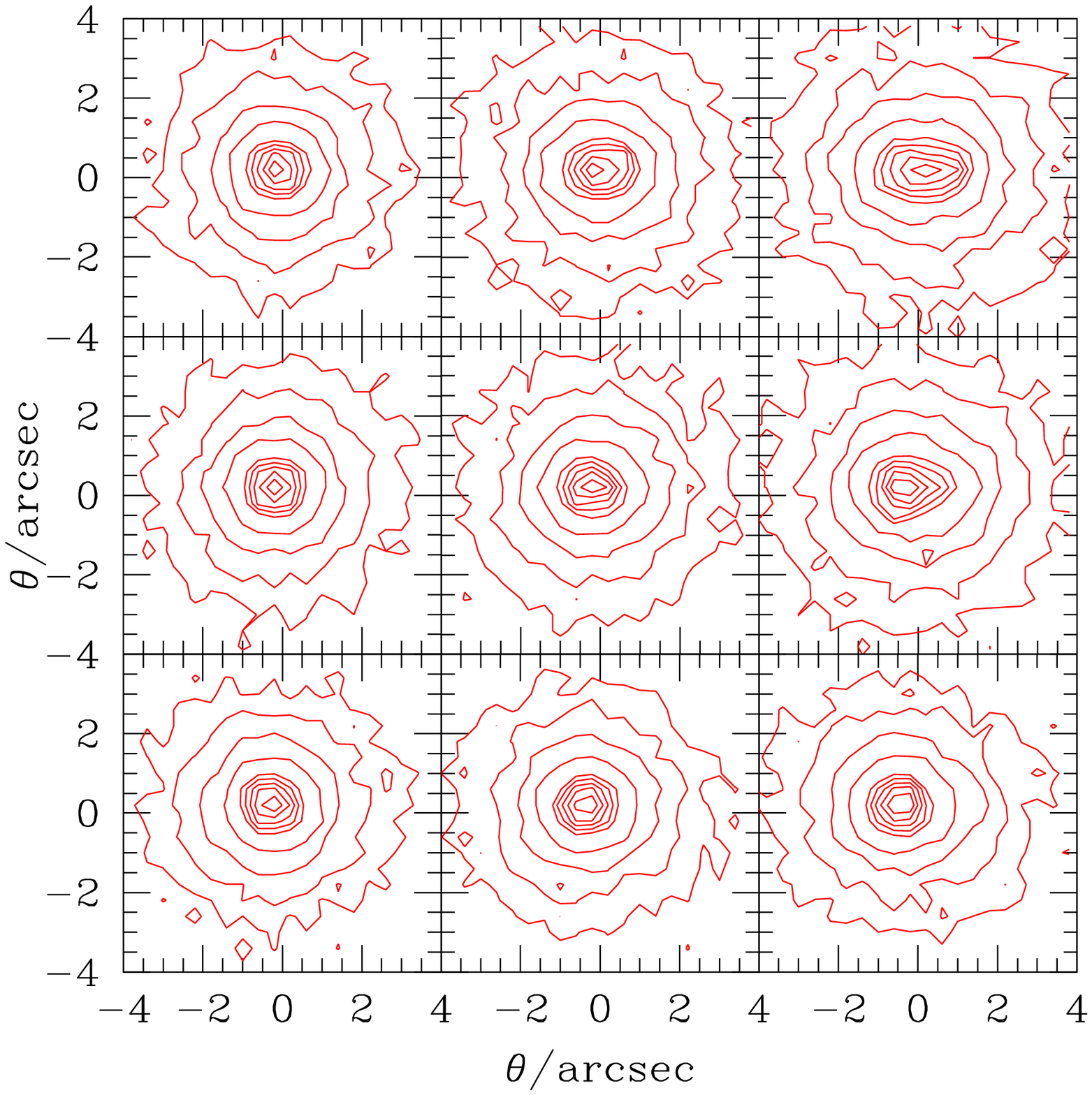}
 \caption{Simulations of two-component images.  Each component is a 
          deVaucoleur profile with half light radius 2\arcsec, 
          the effects of seeing are modeled by a symmetric Gaussian 
          with FWHM = 1.5\arcsec, and the photon counts are binned 
          in square pixels which are 0.4\arcsec on a side.  Together,  
          the two components sum to $r=17.5$~mags.  
          Panels from left to right show what happens as the 
          separation between the objects increases from 0.4 to 
          0.8 to 1.2\arcsec, and from top to bottom show brightness ratios 
          of 1, 2.5 and 10 (the brighter component is on the left).  
          Contours show isophotes which are 
          0.9, 0.8, 0.7, 0.6, 0.5, 0.25, 0.125, 0.0625 and 0.03125 
          times the maximum.  Seeing limits evidence for superposition 
          from the photometry to angular separations $\ga$1\arcsec, 
          and this minimum scale increases if the brightnesses 
          of the two components are very different.  }
 \label{simimages}
\end{figure*}

In principal, the superposed galaxy could be any morphological type and 
any intrinsic luminosity, since the magnitude limit only constrains the 
combined luminosities of the blend.  In practice, we would only really 
notice the superposition if the apparent brightnesses do not differ by 
more than a factor of ten (e.g. Figure~\ref{simimages}).  
Moreover, if the second object showed emission lines, the blend would 
probably have been excluded from the sample.  
Therefore, as a first simple estimate of the $\bar n$ we are after we 
use the total number density of early-types, whatever their luminosity.  
Since we see $\sim 30,000$ galaxies in $\sim 2200$ sq. deg., 
there are about $10^{-6}$ galaxies per square arcsec, 
so the probability of overlap is 
 $\sim (\pi\times 10^{-6}) (\theta/{\rm arcsec})^2$; 
we would expect to find one pair separated by less than 1 arcsec 
for every few hundred thousand galaxies.  
This is significantly smaller than we think we find, presumably because 
by ignoring clustering, we have underestimated the number of close pairs.  

To approximately account for clustering, suppose that the spatial 
correlation function is 
$\xi(z,p) = [r_0^2/(z^2 + p^2)]^{\gamma/2}$, where $z$ and $p$ 
denote comoving distances along and perpendicular to the line of 
sight.  Provided $\gamma>1$, the number of excess pairs at projected 
distance $p$ is 
$w(p) = \int dz\,\xi(z,p) = g(\gamma)\,r_0\,(r_0/p)^{\gamma-1}$
where 
$g(\gamma) = \sqrt{\pi}\Gamma[(\gamma-1)/2]/\Gamma[\gamma/2]$.  
The typical number of extra pairs within projected comoving distance 
$p$ of a randomly chosen galaxy is 
\begin{equation}
 \bar N_{\rm cl}(<p) = \bar n\, 2\pi \int {\rm d}p\, pw(p) 
                     = \bar n\, {2\pi p^2\, w(p)\over 3-\gamma}.  
\end{equation}
For a sample of galaxies which is similar to our early-type sample, 
Budavari et al. (2003) find that $\gamma=2$ and $r_0=8h^{-1}$Mpc 
(their reddest, most luminous sample).  If we approximate 
$p \approx (cz/H)\,\theta$, then 
\begin{equation}
 \bar N_{\rm cl}(<\theta|z) \approx 0.0043
                       \left({\bar n\,{\rm Mpc}^3\over 0.001h^3}\right) 
                       \left({r_0\,h\over 8\,{\rm Mpc}}\right)^2 
                       \left({z\over 0.22}\right) 
                       \left({\theta\over {\rm arcsec}}\right)
\end{equation}
Note that the number of random pairs on these scales, 
 $\sim\pi\times 10^{-6}\,(\theta/{\rm arcsec})^2$, 
is substantially smaller, so can safely be neglected.  

Integrating over the range of observed redshifts will only modify this 
estimate slightly.  For instance, the observed redshift distribution is 
well approximated by 
$dz\,dN/dz \propto 2\,(dz/z)\,(z/z_m)^3 \exp[-(z/z_m)^2]/\Gamma(3/2)$, 
with $z_m = 0.11$, so 
\begin{equation}
 \int dz\,{dN\over dz}\,\bar N_{\rm cl}(<\theta|z) =
   {2\over\sqrt{\pi}}\,\bar N_{\rm cl}(<\theta|z_m).
\end{equation}
Thus, this analysis suggests that about one in every four hundred 
objects will be a blend.  Notice also that, at least on small scales, 
the number of blends increases as $\theta^{3-\gamma}$; for $\gamma = 2$, 
the number scales linearly with the allowed angular separation between 
the blends.  

This analysis assumes that $\xi$ is a power-law all the way down to 
vanishingly small scales.  If $\xi$ is shallower on small scales, then 
the predicted number of superpositions is smaller, and scales 
approximately as $\theta^{3-\gamma_{\rm eff}}$, where $\gamma_{\rm eff}$ 
is the effective slope on scales of order 10 kpc (recall that 1\arcsec\ 
at $z=0.3$ is 4.4 kpc).  

To account more carefully for the effects of clustering and the 
magnitude limit is more complicated.  We have chosen to do so by 
performing Monte-Carlo simulations as follows.  Note that this 
analysis is not intended to be definitive---we only wish to 
demonstrate how more sophisticated models of the early-type galaxy 
distribution can yield substantially more information about the 
likelihood of superposition.  Conversely, the fraction of 
superpositions constrain the various parameters of such Monte-Carlo 
models.  

We start with the Very Large Simulation (VLS; 
Yoshida, Sheth \& Diaferio 2001) of a $\Lambda$CDM cosmology.  
This simulation followed the evolution of $512^3$ particles in a 
$479h^{-1}$Mpc cube; the particle mass in the simulation is 
$6.86\times 10^{10}h^{-1}M_\odot$, so halos more massive than 
$\sim 1.3\times 10^{12}h^{-1}M_\odot$ are reasonably well represented 
by the simulation.  
A catalog of the positions and masses of the dark matter halos in this 
simulation has been made available by the Virgo consortium 
({\tt http://www.mpa-garching.mpg.de/Virgo}).  
We model the SDSS early-type galaxy distribution by populating the 
simulated halos with model galaxies in such a way that the 
luminosity function, and the luminosity dependence of clustering 
are approximately reproduced.  In particular, Zehavi et al. (2005) 
studied the clustering of galaxies as a function of luminosity in 
the SDSS.  Although they did not study clustering as a function of 
morphological type, we use some aspects of their results to guide 
the construction of our mock catalogs as follows.  

We assume that the mean number of early-type galaxies increases with 
halo mass as  
$\langle N_{\rm gal}|M_{\rm halo}\rangle = 1 + M_{\rm halo}/23M_{\rm min}$ 
provided $M_{\rm halo}>M_{\rm min}$, and it is zero in lower mass haloes.  
The parameter $M_{\rm min}$ is chosen to match the number density of 
early-type galaxies reported by Bernardi et al. (2003b):  
in our case, $M_{\rm min} = 1.5\times 10^{12}h^{-1}M_\odot$.  
With this prescription, about twenty percent of the galaxies are in 
halos which host at least one other galaxy.  In all halos more 
massive than $M_{\rm min}$, we place the first galaxy at the halo centre, 
then draw a Poisson random number $N_{\rm sat}$ which has mean 
$M_{\rm halo}/23M_{\rm min}$ (a Poisson distribution for the satellite 
galaxies is motivated by the work of Kravtsov et al. 2004), 
and distribute the $N_{\rm sat}$ satellite galaxies around the halo 
centre so the resulting density run of galaxies resembles an 
NFW profile (Navarro et al. 1997).  

To insure that the mock galaxy sample has the correct distribution 
of luminosities, we generate the Lognormal distribution which 
Bernardi et al. (2003b) find describes this early-type sample well.  
The results of Zehavi et al. (2005) suggest that the central galaxy 
in a halo is almost always substantially more luminous than the others, 
and that its luminosity increases monotonically with the mass of its 
host halo.  In contrast, the distribution of satellite galaxy 
luminosities depends only weakly on parent halo mass.  To incorporate 
such an effect into our mocks, we rank order the luminosities and 
the host halo masses, assign the brightest luminosity to the galaxy 
at the centre of the most massive halo,  and work our way down the set 
of luminosities and halos, assigning successively fainter luminosities 
to central galaxies of host halos with ever lower masses.  
Once all central galaxies have been assigned luminosities, the 
fainter luminosities which remain are assigned to satellites without 
regard to the masses of the parent halos.  

We then model the SDSS survey as a cone oriented in a random direction 
within the box, and compute angular positions, redshifts and apparent 
magnitudes for each simulated `galaxy'.  
In this way, we can simulate the chance of having a blend which would 
have been bright enough to satisfy the SDSS magnitude limit.  

The simulation box is $L=479h^{-1}$Mpc comoving on a side, which means 
that observations out to $z \sim 0.16$ are straightforward to model.  
Since we would like to reach to redshifts of order 0.3, we surround the 
initial box by copies of itself, each rotated by a random angle, to avoid 
spurious projection effects.  This is not ideal, particularly because 
the VLS halo catalogs are taken from a single snapshot at $z=0$, so 
our mock catalogs do not account for evolution along the lightcone.  
A similar analysis using the Hubble Volume lightcone outputs is only 
possible for halos more massive than $5\times 10^{13}h^{-1}M_\odot$.  
For these more massive halos, the VLS and Hubble Volume catalogs 
yield similar results, suggesting that evolution along the lightcone 
is not an important effect.  
We also constructed mock catalogs with lightcone effects built-in 
using the {\tt PTHalos} algorithm (Scoccimarro \& Sheth 2002).  
These allow us to probe smaller masses;  they too suggest that 
the single snap-shot VLS simulations are sufficiently accurate for 
our purpose.  


In our simulations, about one out of every five hundred objects with 
apparent magnitudes in the range $14.5\le m_r\le17.75$ is a 
superposition of two galaxies which are separated by less than one 
arcsec.  (If we also impose a cut on the ratio of the apparent 
brightnesses of the two components, to model the fact that if the 
smaller component contributes less than ten percent of the light we 
are unlikely to notice it, then the predicted number of identifiable 
superpositions falls slightly.)  On arcsecond scales, the fraction 
increases slightly faster than linearly with increasing angular 
separation.  (Thus, our analytic clustering estimate was not far-off.)  
Of these, only about ten percent come from objects which are in different 
halos; for the majority of pairs, both galaxies are in the same cluster.  
The spectra shown in Figure~\ref{doubleimgspec} suggest that separations 
in velocity space are typically between 500 and 1000~km~s$^{-1}$, 
although it is difficult to identify superpositions from the spectra alone 
if the redshift differences are smaller than $\sim$500~km~s$^{-1}$.  
If these line-of-sight separations are due to virial motions, then they 
correspond to virial masses of order $5\times 10^{14}h^{-1}M_\odot$, 
consistent with the simulations (recall that only halos with mass 
greater than $\sim 4\times 10^{13}h^{-1}M_\odot$ contain more than 
one early-type galaxy, so we expect superpositions from galaxies in 
substantially more massive halos).  



\subsection{A.2 Evidence for superpositions}\label{evidence}
We use a combination of photometric and spectroscopic information to 
determine if an object is likely to be a superposition, and we then 
use a combination of two methods for estimating the velocity dispersions 
of the two components (although the S/N ratios of the spectra are 
relatively low, so these estimates are crude).  
If the isophotes of the image are asymmetric we flag the object as 
a possible blend.  
Figure~\ref{simimages} illustrates that evidence for superposition 
from the photometry is limited to angular separations larger than 
$\sim$1\arcsec, and brightness ratios of order 10. 
Therefore, we also use the cross-correlation method (e.g. Simkin 1974; 
Tonry \& Davis 1979) as a simple diagnostic:  
we flag the object as a possible blend if the highest peak of the 
cross-correlation function is asymmetric.  
We then re-fit for the velocity dispersion, this time allowing for the 
possibility that the observed spectrum contains light from two sources.  
Our method is described in the next subsection.  

We label as doubles all objects whose spectra are significantly better 
fit by two components than one; this was true for about half the 
objects.  For the other half, the spectra and the photometry gave 
ambiguous results (e.g., neither the image nor the cross-correlation 
peak showed significant asymmetry, or the best two-component fit returned 
the same template with negligible redshift separation (i.e., smaller than 
one pixel).  Although a substantial fraction of these objects may well 
be massive single galaxies, follow-up observations are required to 
produce conclusive evidence against the superposition hypothesis.  


The next subsection describes our analysis of the spectra, the 
results of which are shown in Figures~\ref{uniqueimgspec} and 
Figure~\ref{doubleimgspec}.  Note that in many instances, the spectra 
show more compelling evidence for superposition than do the images.  
Note also that a significant fraction of superpositions are in fields  
which are not particularly crowded.  
(Only a few representative examples are presented here; 
the electronic version of this article shows similar figures for all 
the objects in our sample.)

\subsection{A.3 Velocity dispersions}\label{veldisps}
The velocity dispersion is usually estimated by starting with a high 
signal-to-noise template spectrum, and then finding that function 
which, when convolved with the template, yields the closest match 
to the observed spectrum.  Fourier (Simkin 1974; Sargent et al. 1977; 
Tonry \& Davis 1979) and real-space techniques 
(Franx, Illingworth \& Heckman 1989; Rix \& White 1992) have been developed 
for doing this.  
The accuracy of the estimated $\sigma$ depends crucially on judicious 
choice of template; if the template is a poor match to the object of 
interest (e.g., using a late-type stellar template to fit the spectrum 
of an early-type galaxy), then all the techniques above will yield a 
biased answer.  

To illustrate, the top pair in each set of panels in  
Figure~\ref{doubleimgspec} show the results of the cross-correlation 
and the direct-fit methods.  
In many cases, the cross-correlation function shows two distinct 
peaks, indicating the spectrum is almost certainly a superposition 
of two objects.  The curves show the result of fitting the sum of two 
Gaussians to the cross-correlation function.  The separation between 
the peaks is an estimate of the separation in velocity space between 
the two components, the widths of the two Gaussians yield estimates of 
the two velocity dispersions, and the relative amplitudes of the 
normalized Gaussians yield estimates of the relative apparent 
brightnesses of the components.  The separations are almost always 
less than 1000~km~s$^{-1}$.  

The direct fit method used by the SDSS assumes that the observed 
spectrum is the convolution of a template spectrum with a single 
Gaussian.  Often, it shows that $\chi^2$ of the difference between 
the observed spectrum and the broadened template has a very well-defined 
minimum, even though the cross-correlation method clearly shows the 
presence of two peaks.  Evidently, by assuming an incorrect broadening 
function (a single Gaussian) the direct fitting method can yield 
misleading results.  On the other hand, simulations show that, if 
either the signal-to-noise ratio is small, or the template is really a 
poor match to the observed spectrum, then the peak of the 
cross-correlation function can become asymmetric.  In the cases where 
the two peaks are well-separated, this is not a concern, but there are 
several other cases in which the separation is small enough that the 
evidence for two components is less compelling.  In such cases, should 
we interpret a deep and narrow minimum from the direct-fit method as 
indicating that the object is, in fact, a single?  
To address such cases, we have modified the direct-fitting method as 
follows.  

The main weakness of the direct fit method was the assumption that 
the broadening function was a single Gaussian, or, more specifically, 
that the broadened template is actually a good description of the 
observed spectrum.  
Most stellar and globular cluster based templates are built from the 
spectra of relatively nearby objects.  Therefore, there are few available 
templates which are suitable for matching the spectra of massive 
early-type galaxies, particularly those which have both super-solar 
metallicities and $\alpha$-element abundance ratios.  This is a 
particular concern, because chemical abundances are expected to 
correlate with velocity dispersion, so one might worry that templates 
constructed from local stellar populations will give increasingly 
biased answers for the objects of most interest to us---the most 
massive early-type galaxies.  

Recently, Bernardi et al. (2003d) have compiled a large catalog of 
early-types, from which they constructed composite spectra of high 
signal-to-noise ratio (S/N$\sim 80$).  It is these composite spectra 
which we use as our templates because, in principle, they already 
incorporate the effects on the spectrum of changing chemical abundances 
with increasing velocity dispersion.  

We construct a library of composites, shifted by various amounts 
(between $-0.004\le\Delta z\le 0.004$ in steps of $\Delta z=0.0001$)
with respect to one another (i.e., the steps are approximately twice 
the size of a pixel in the SDSS spectrograph).  We then find that pair 
of shifted composites which most closely match (in a $\chi^2$ sense) 
the observed spectrum in (log) wavelength space.  
Solution of this minimization problem requires inversion of an 
$m\times m$ matrix where $m$ is the product of the total number of 
composites and the total number of shifts.  In this respect, our method 
is essentially that of Rix \& White (1992), except that, because our 
template spectra are already broadened, the `broadening function' to be 
found is the separation between the two composites.  We treat the overall 
normalization of each composite template as a free parameter, so that 
the fitting procedure also returns the fraction of the total light 
in each component.  

The best-fit spectra returned by this method (smooth red lines) are 
compared with the observed spectra (noisier black lines) in the sets 
of three panels shown in Figure~\ref{doubleimgspec}.  
Rather than showing the entire wavelength range, we have chosen to 
highlight the regions around the H and K lines, the $g$ band, and the 
Mg doublet.  The text indicates the 
velocity dispersions and redshift separations of the two composites, 
and the fraction $f$ of light contributed by the first composite.  
For comparison, the estimated velocity dispersion based on fitting a 
single broadened Gaussian is shown in the upper left hand corner; 
the associated spectrum is shown in blue (dotted).  
In most cases, the two-component fit is a significant improvement.  
It is also instructive to compare the estimated separations yielded 
by this method with the cruder estimates based on the cross-correlation 
method (cruder because the cross-correlation method compares a single 
small $\sigma$ template with the observed galaxy spectrum, rather than 
a pair of composite spectra). 

\section{B. Images and spectra of objects with large velocity dispersions}
Figure~\ref{uniqueimgspec} shows images and spectra of objects 
which are not obvious superpositions.  
Figure~\ref{doubleimgspec} shows the corresponding information 
for objects which are superpositions.  
In all cases, top left in each series of panels shows fields approximately 
$1\arcmin\times 1.5\arcmin$ and $7.6\arcsec\times 10.4\arcsec$ in size 
centred on the objects (each pixel is 0.4\arcsec\ on a side).  
The SDSS spectrograph fibers are each 3\arcsec\ in diameter, so 
neighbours more distant than this are unlikely to affect the observed 
spectrum.  The top center and top right panels show the results of the 
cross-correlation function and the direct-fit estimates of the 
velocity dispersion.  Bottom panels show sections of the spectrum 
with our best-fitting composite spectra superimposed.  
In Figure~\ref{doubleimgspec} (objects classified as superpositions), 
the best two-component fit is also shown.  Two-component fits are also 
shown in some cases of Figure~\ref{uniqueimgspec} where the evidence 
for two components is reasonable but not compelling.  

The electronic version of this article shows similar figures for all 
the objects in our sample; only a few representative examples are 
presented here.  These examples were chosen to illustrate which 
features in the images or spectra help determine whether or not the 
object in question is a superposition.  The first three show objects 
we classified as unlikely to be superpositions, whereas the final three 
are almost certainly superpositions.  

\begin{itemize}
 \item{\it SDSS J151741.7-004217.6}:  This object is in a reasonably 
crowded field, but the surface brightness contours show no clear 
evidence of irregularities.  The spectrum has relatively high S/N, 
and also shows no convincing evidence of superposition:  the cross 
correlation function is symmetric, the minimum of $\chi^2$ for the 
direct fit method is narrow and well-defined, and a single component 
fit provides a good description of the various line-profiles.  

\item{\it SDSS J154017.3+430024.5}:  This object is in a considerably 
more crowded field, and the isophotes in the center are slightly 
irregular.  However, the cross-correlation and direct-fit techniques 
are still relatively symmetric.  Notice that the spectrum has slightly 
lower S/N.  

\item{\it SDSS J141922.4+011457.8}:  This object is relatively 
isolated and the isophotes show no clear evidence of irregularities.  
While the cross-correlation function is not symmetric, the 
asymmetry is not strong enough to provide compelling evidence for 
two components.  

\item{\it SDSS J014157.5-010626.3}:  This object is in a crowded 
field, the isophotes show evidence for two components.
The cross-correlation function suggests slight evidence for two components  
and the minimum of $\chi^2$ from the direct fit method is broad and 
asymmetric.  However, two component fits to the various line-profiles are 
not significantly better than single component fits.  Nevertheless, the 
spectrum is beginning to show rapid oscillations which are not seen in 
single-component spectra.  

\item{\it SDSS J162255.0+455514.7}:  This object is also in a 
crowded field, and the isophotes, the cross-correlation function 
and the direct-fitting method all show evidence for two components.  
Two-component models provide a reasonable description of the 
oscillations in the spectra.  

\item{\it SDSS J080234.9+362100.9}:  This object is in a crowded 
field, and the spectrum shows clear evidence for two components, 
even though the isophotes do not.  Notice again how the two 
component model provides a significantly better description of the 
oscillations in the spectrum.  
\end{itemize}

\setcounter{figure}{0}

\begin{figure*}
 \centering
 \epsfxsize=0.45\hsize\epsffile{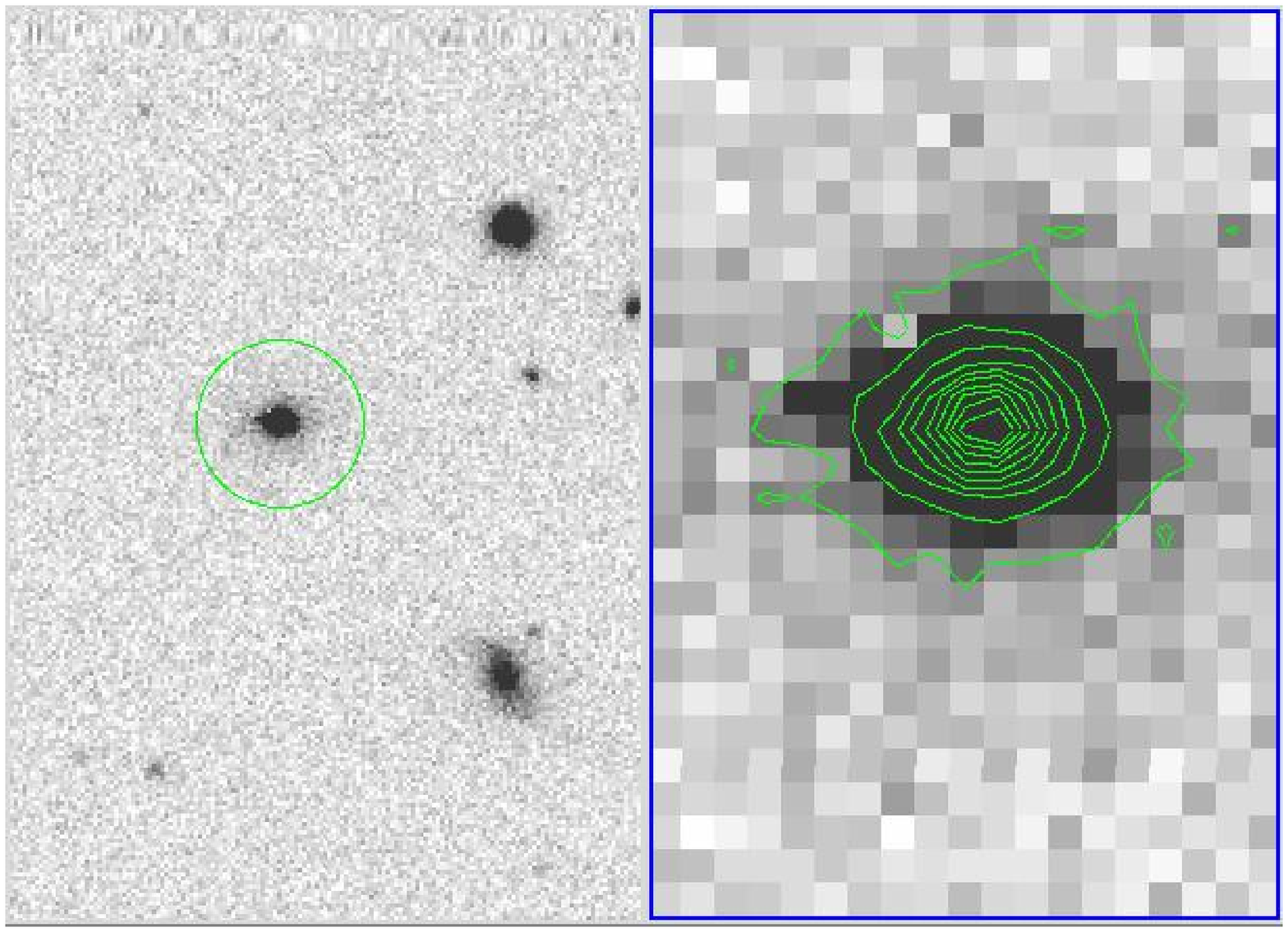}
 \epsfxsize=0.3\hsize\epsffile{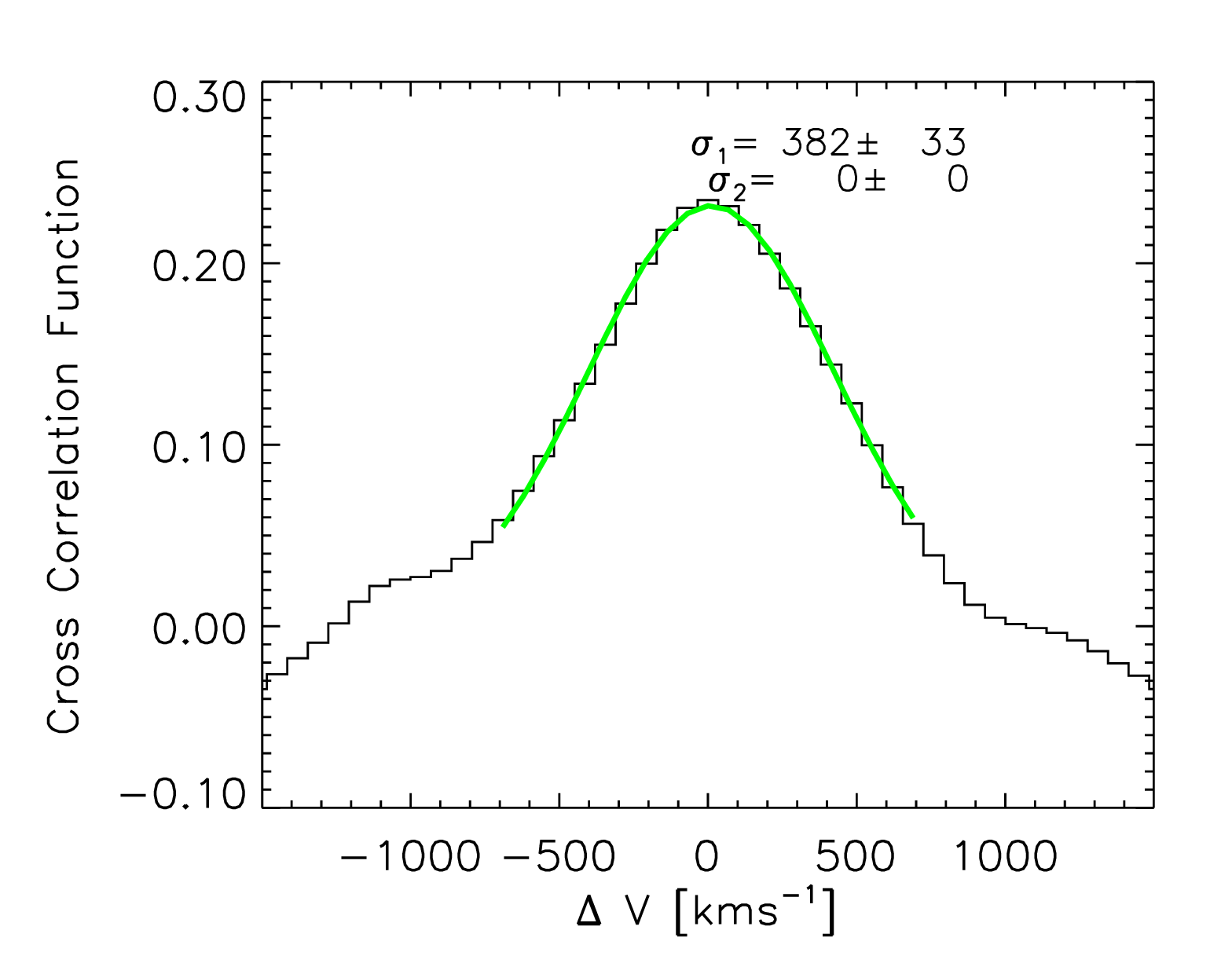}
 \epsfxsize=0.3\hsize\epsffile{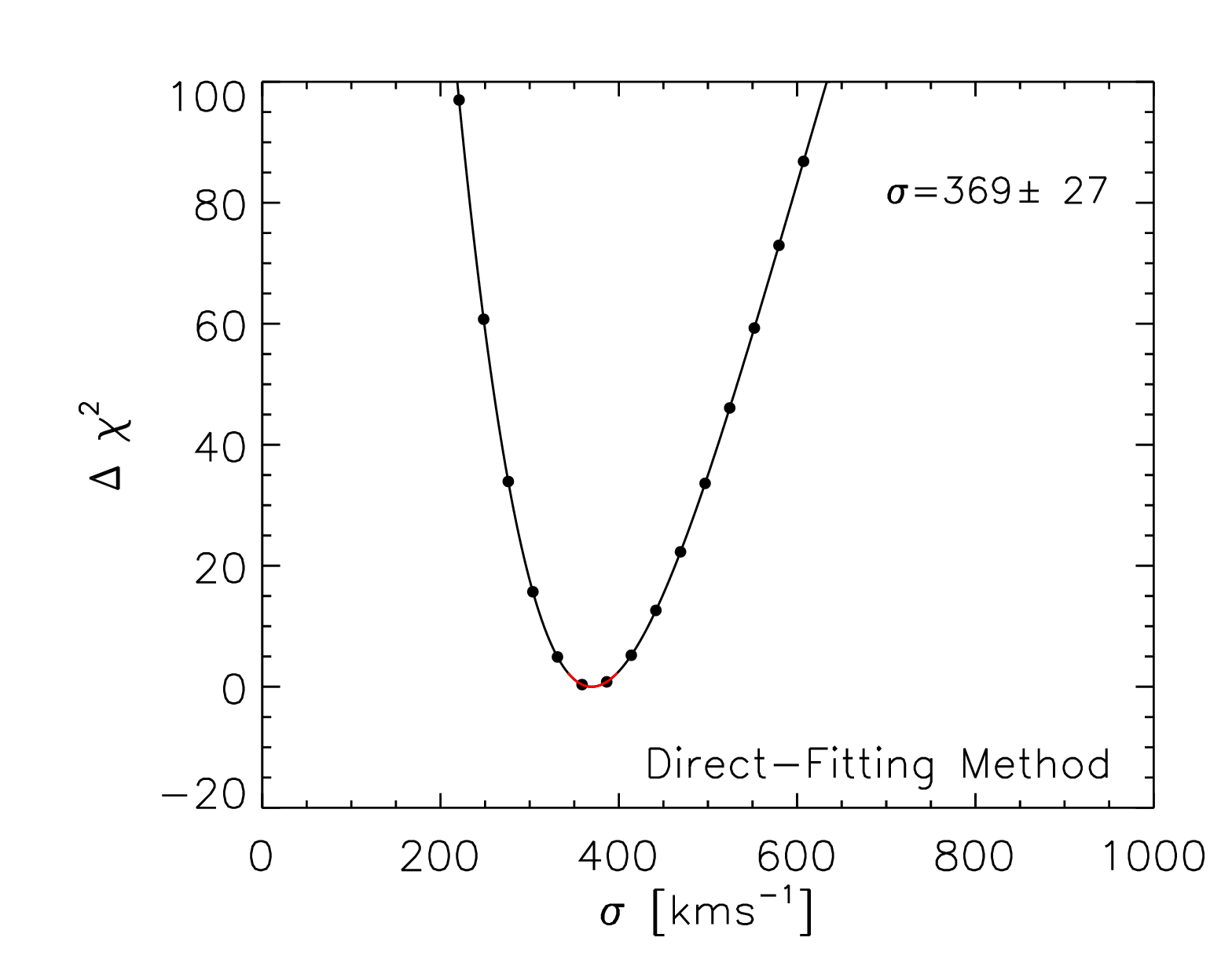}
 \epsfxsize=0.7\hsize\epsffile{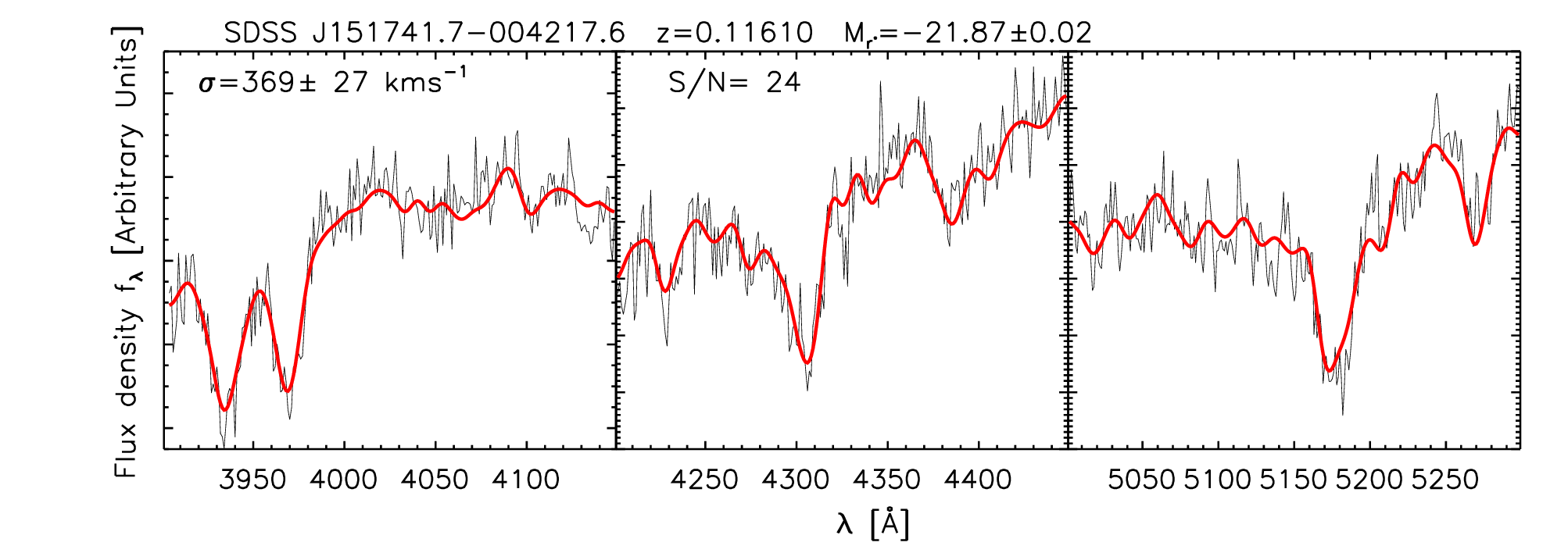}
 \caption{Photometric and spectroscopic properties of objects 
          with $\sigma>350$~km~s$^{-1}$ and S/N$>10$ which are not 
          obvious superpositions.  
          Top left in each series of panels shows fields approximately 
          $1\arcmin\times 1.5\arcmin$ and $7.6\arcsec\times 10.4\arcsec$ 
          in size centred on the objects (each pixel is 0.4\arcsec\ on a 
          side).  
          The SDSS spectrograph fibers are each 3\arcsec\ in diameter, 
          so neighbours more distant than this are unlikely to affect the 
          observed spectrum.  
          The top center panels show the cross-correlation between a 
          template and the observed spectrum.  Double-Gaussian fits to the 
          cross correlation function, shown as two smooth red curves 
          which sum to give the green curve, and yield estimates for the 
          velocity dispersions of the individual components.
          The value of $\chi^2$ around its minimum, computed using a 
          the direct-fit method assuming only a single broadening 
          function, is shown in the top right panel of each series; 
          text shows the estimated velocity dispersion.  
          The bottom panels in each series show 
          the result of using the direct-fit method to determine the 
          pair of composite spectra which best-fit the observed spectrum; 
          red solid lines show the combined spectrum of the best-fitting pair, 
          the parameters of which are given in the left-most panel.  
          For comparison, blue dotted lines show the best-fitting single 
          component spectrum; text at the top of the panel shows the 
          associated estimate of $\sigma$. In some cases the spectrum
          is best-fitted just by a single component; in these cases
          only a red solid line, showing the best-fitting single 
          component, is presented.
          (The values of $\sigma$ reported in this Figure are not aperture 
           corrected.)}
 \label{uniqueimgspec}
\end{figure*}

\setcounter{figure}{0}

\begin{figure*}
 \centering
 \epsfxsize=0.6\hsize\epsffile{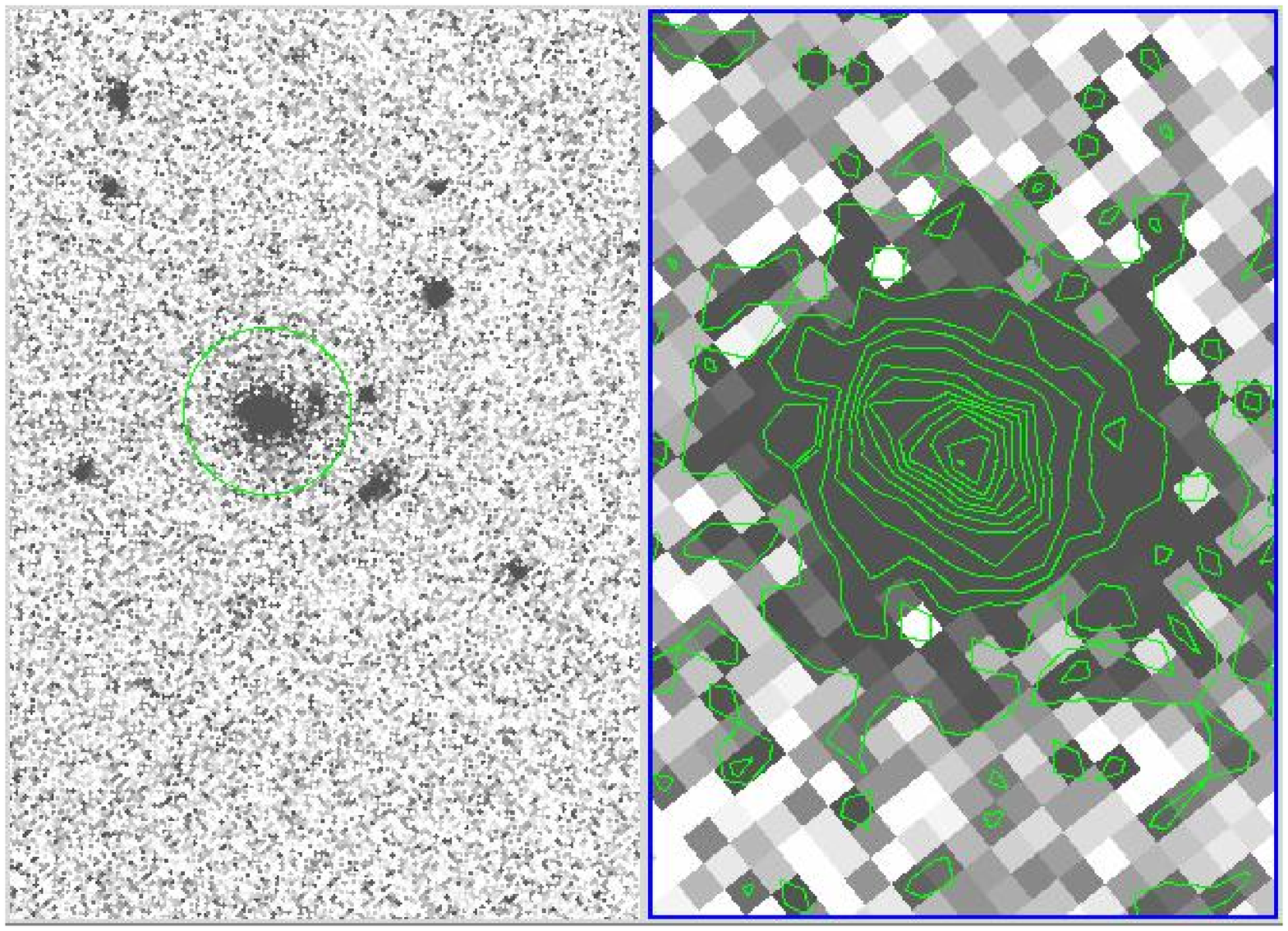}
 \epsfxsize=0.35\hsize\epsffile{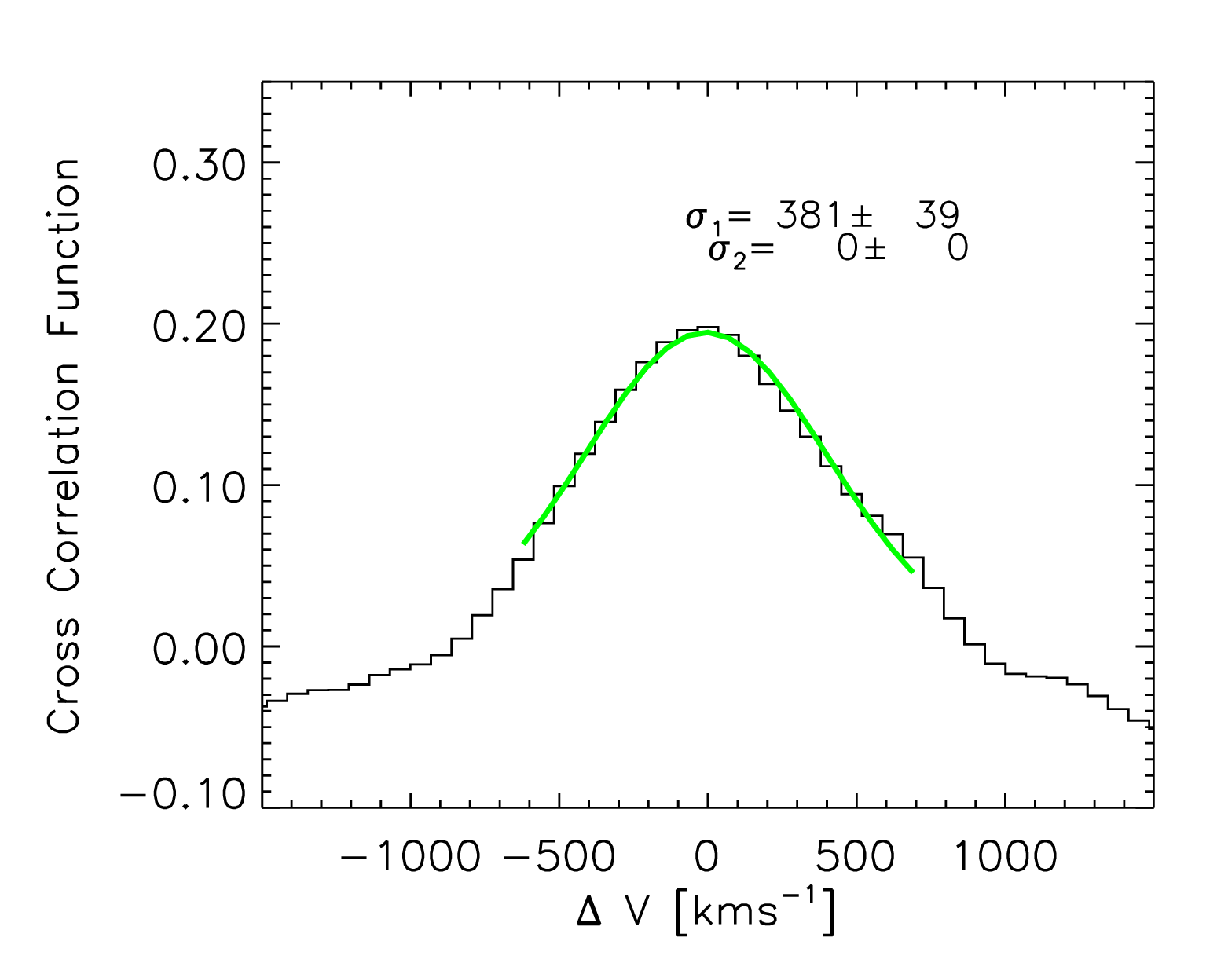}
 \epsfxsize=0.35\hsize\epsffile{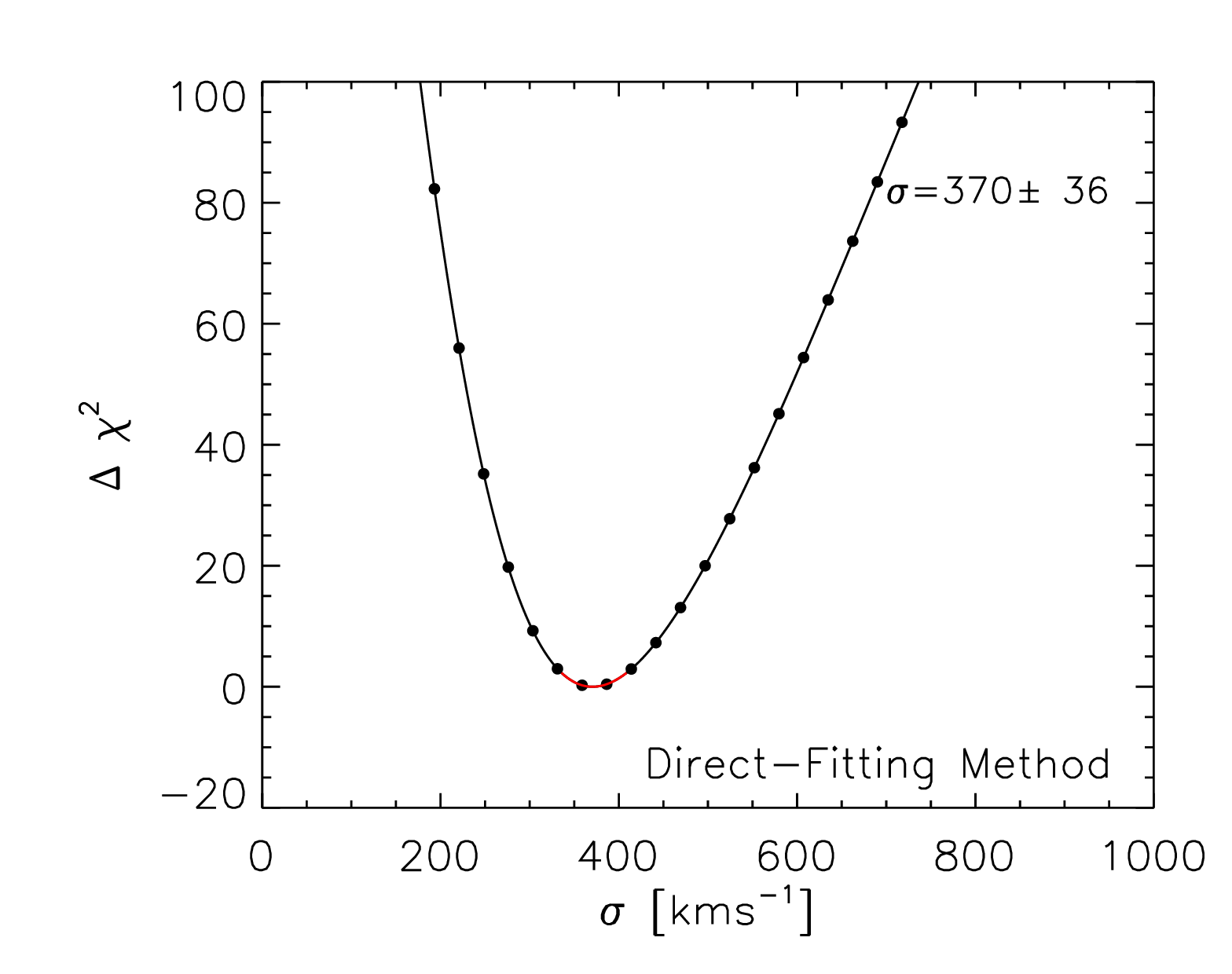}
 \epsfxsize=0.8\hsize\epsffile{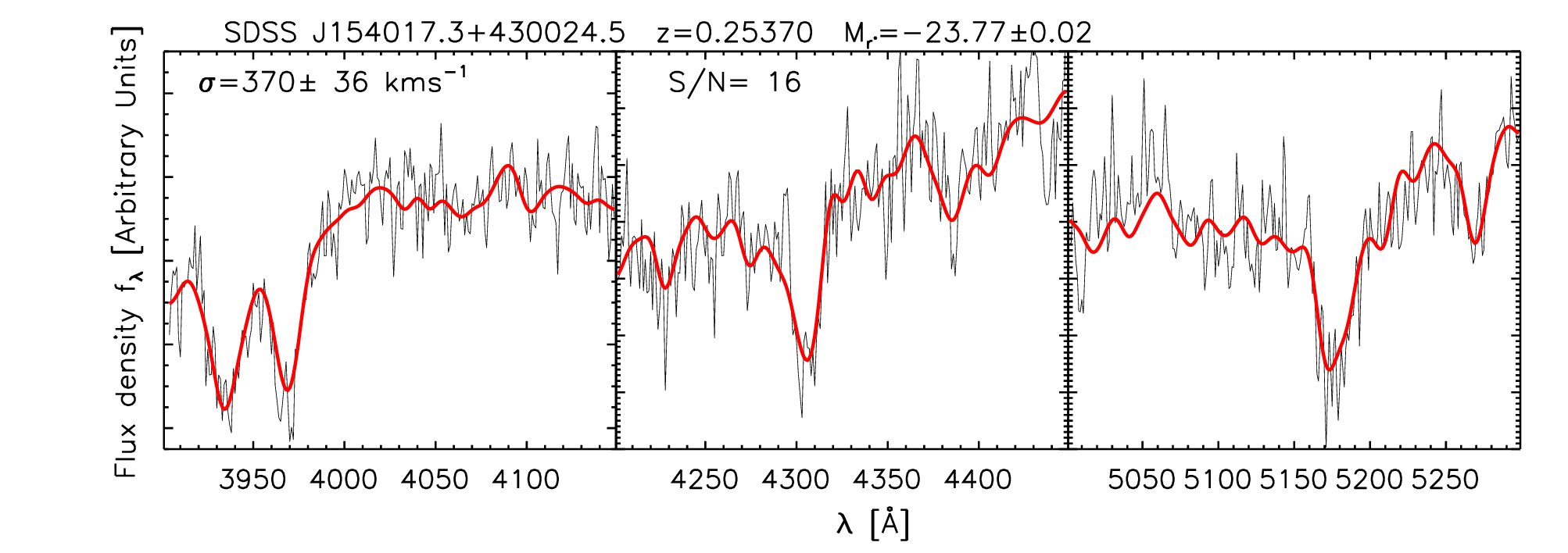}
 \caption{Continued.}
\end{figure*} 

\setcounter{figure}{0}

\begin{figure*}
 \centering
 \epsfxsize=0.6\hsize\epsffile{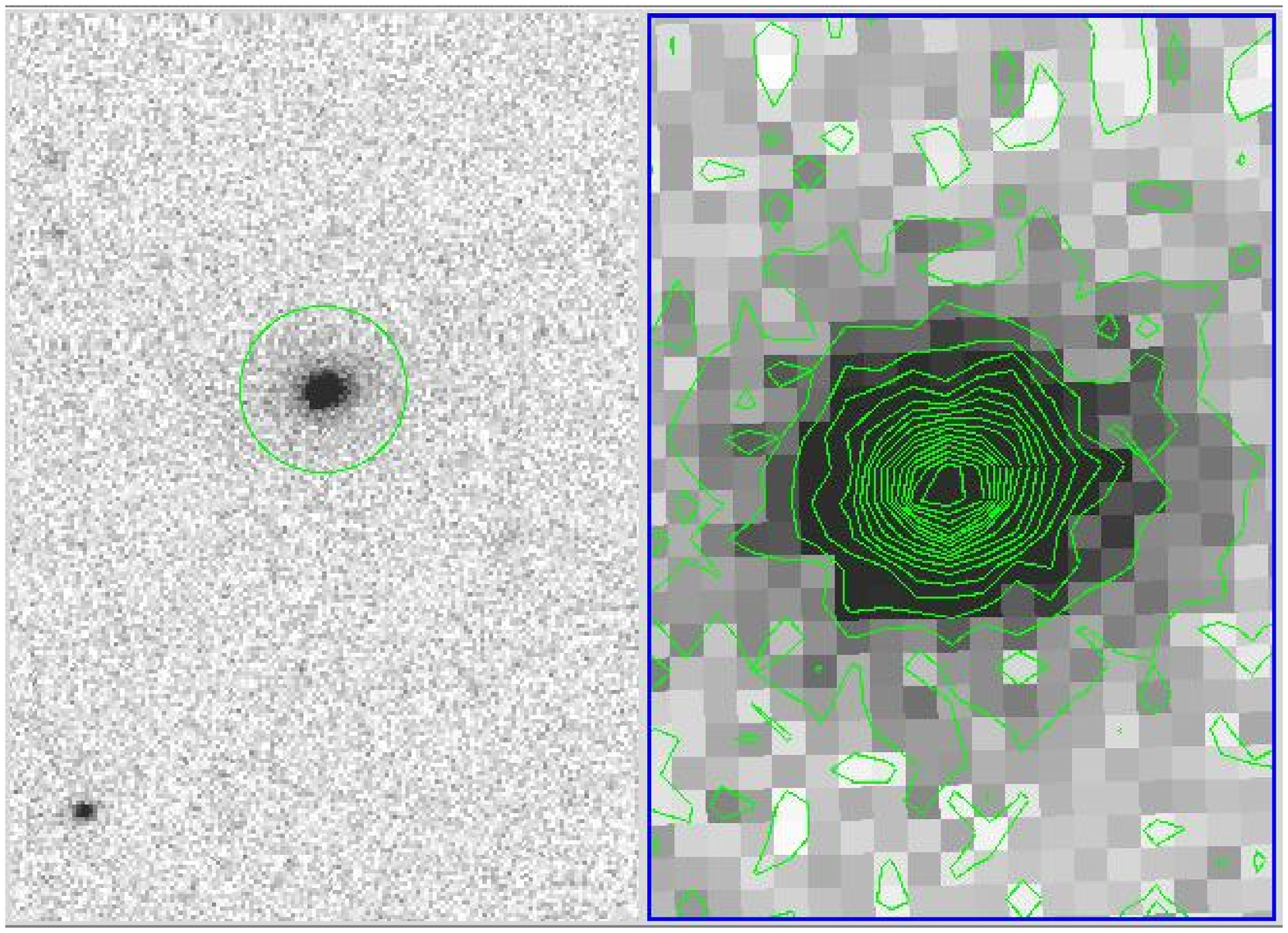}
 \epsfxsize=0.35\hsize\epsffile{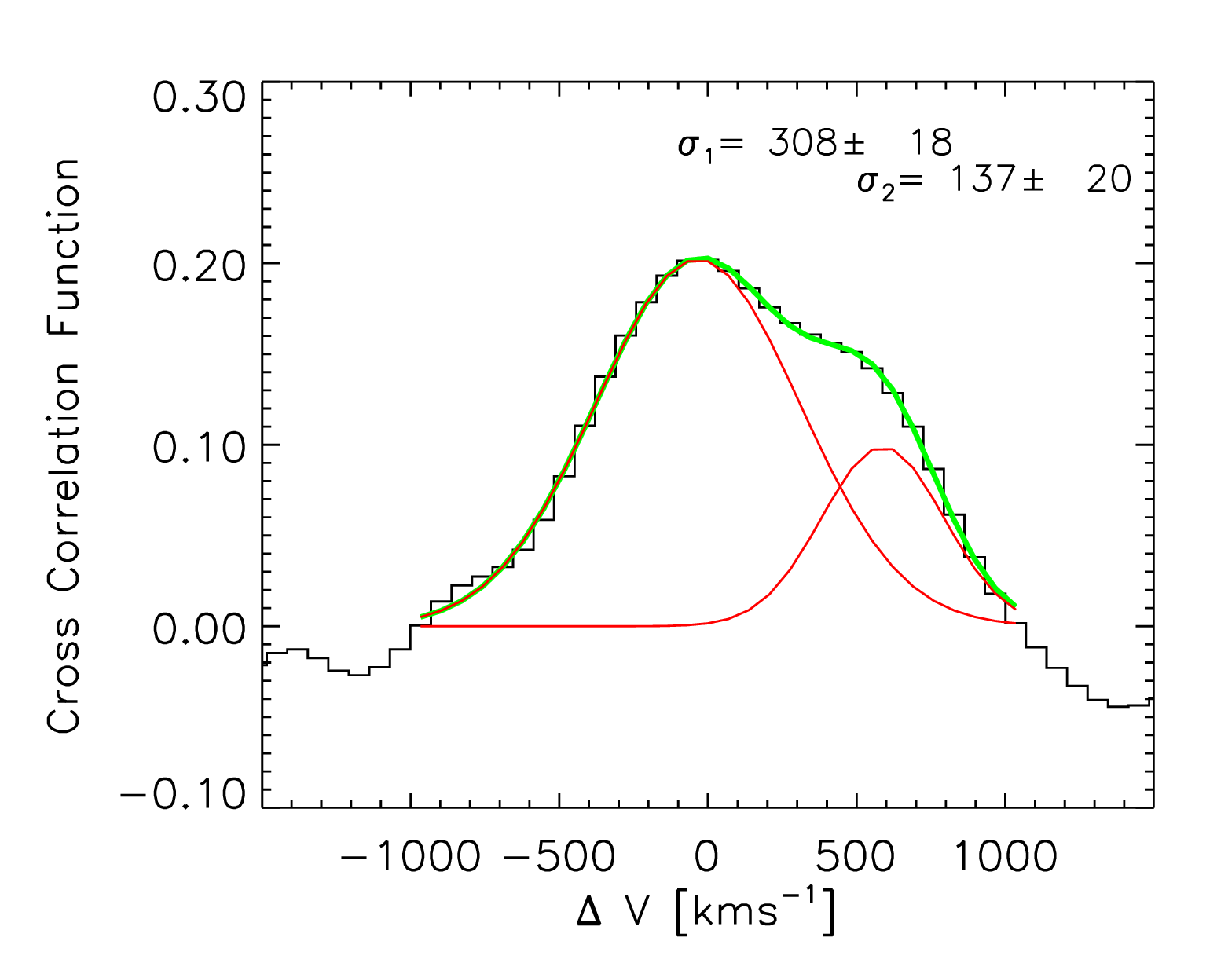}
 \epsfxsize=0.35\hsize\epsffile{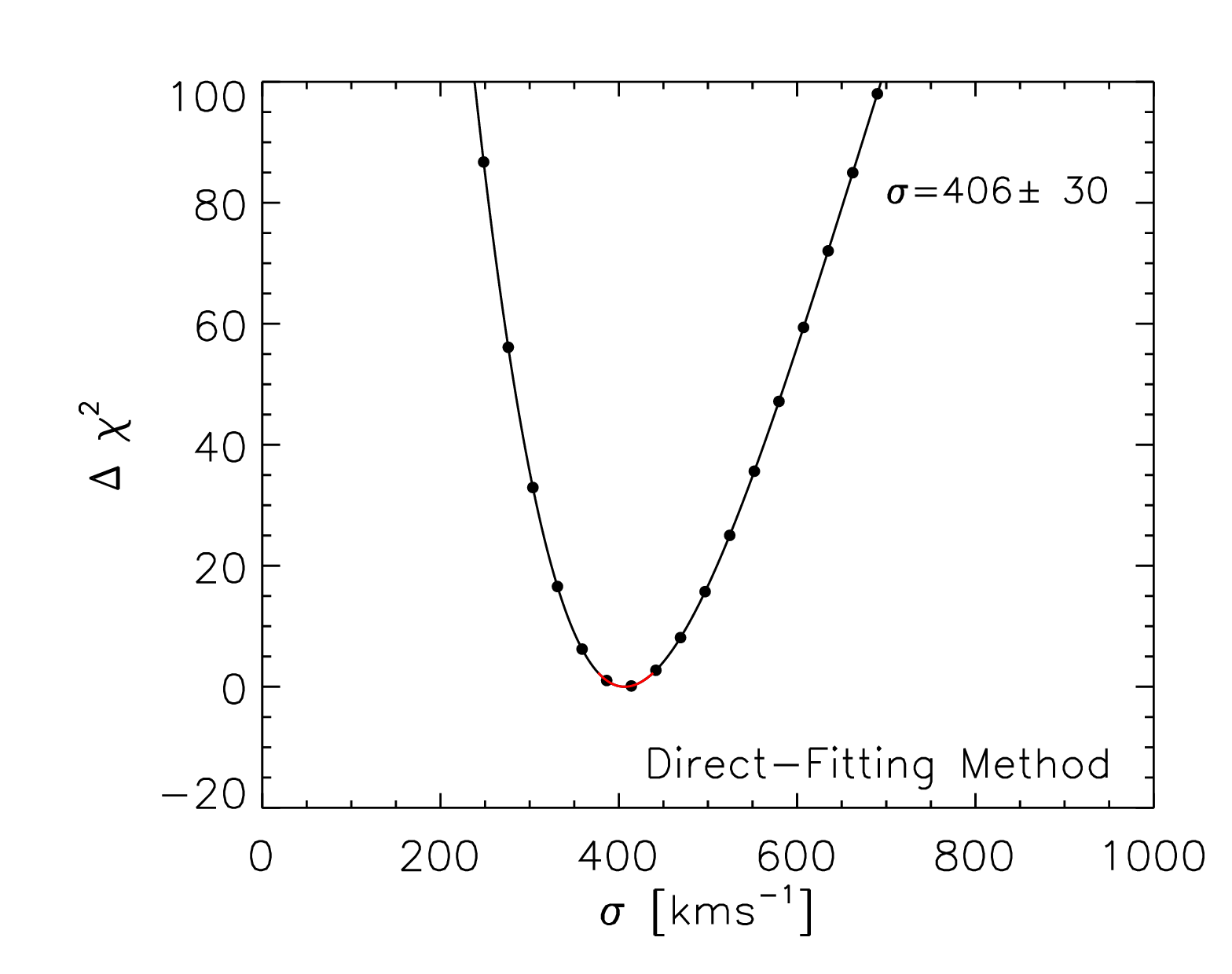}
 \epsfxsize=0.8\hsize\epsffile{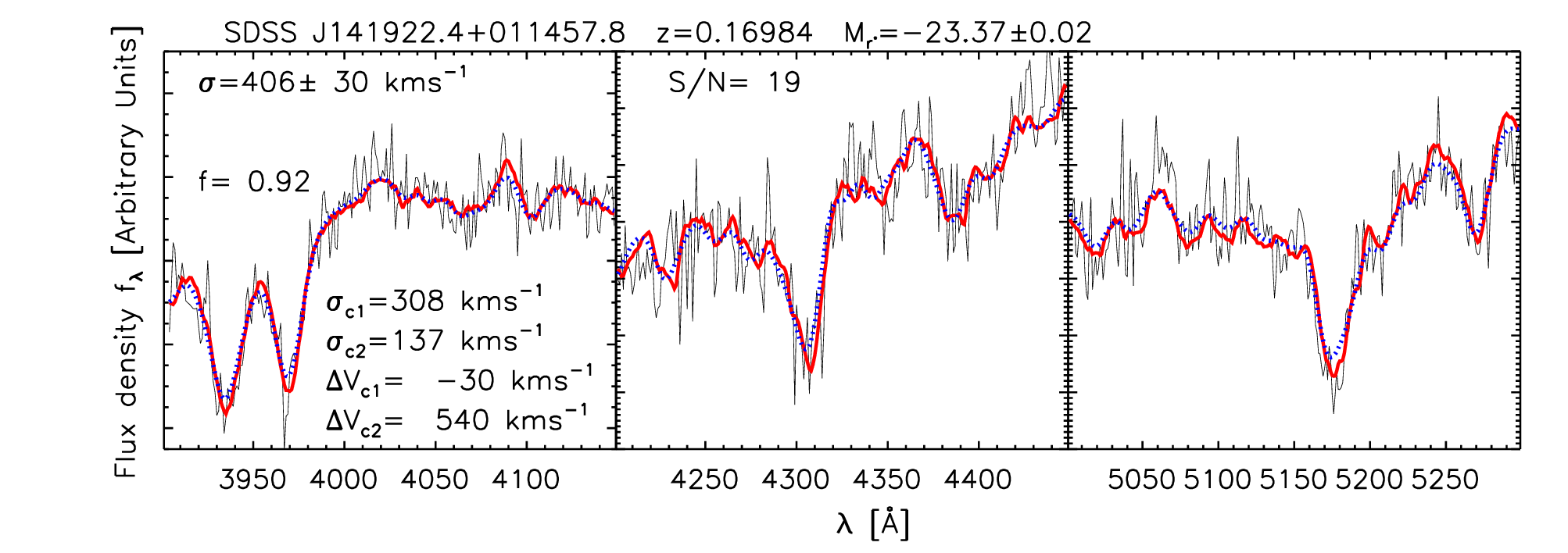}
 \caption{Continued.}
\end{figure*}

\clearpage


\setcounter{figure}{1}

\begin{figure*}
 \centering 
 \epsfxsize=0.6\hsize\epsffile{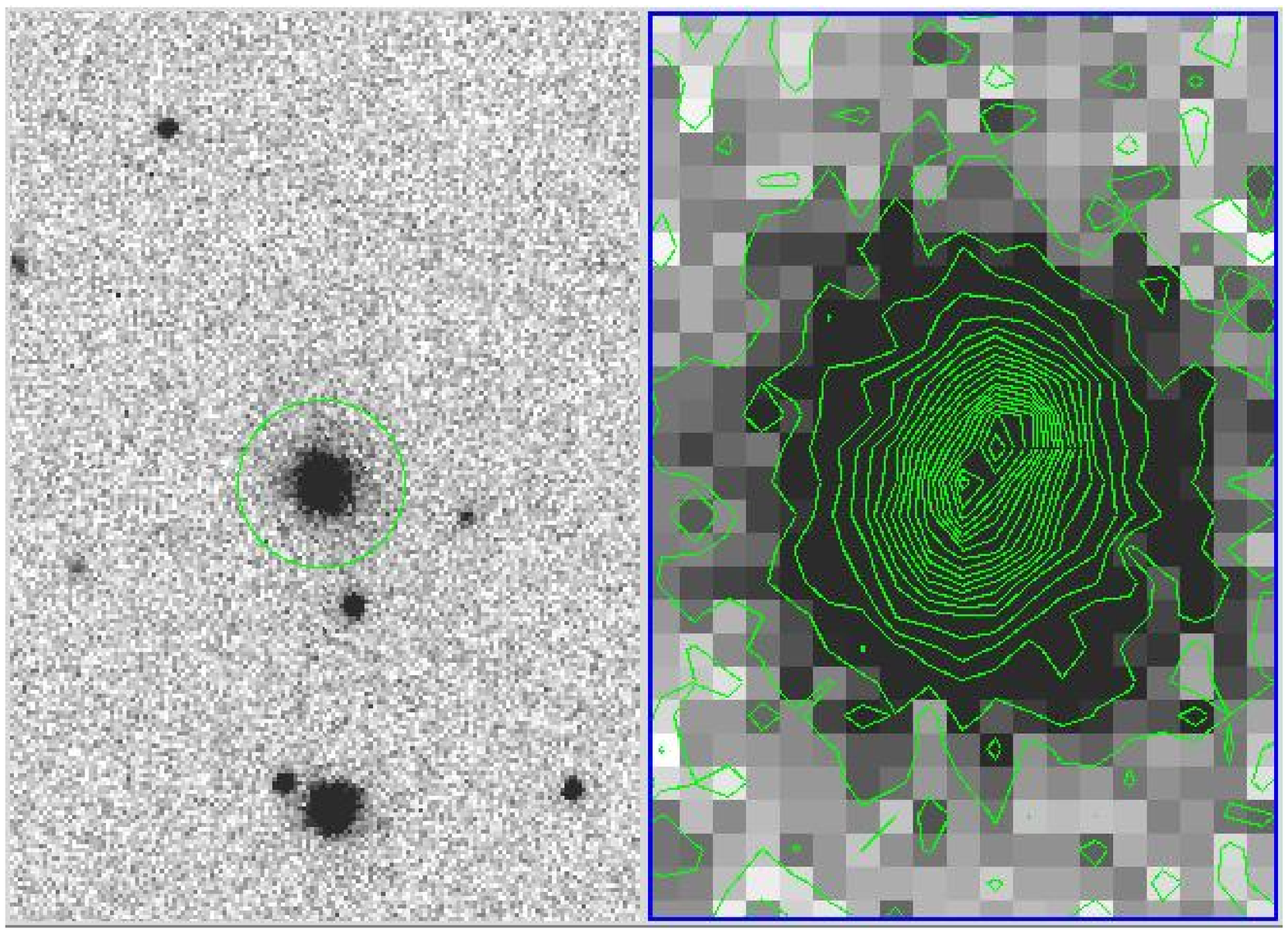}
 \epsfxsize=0.35\hsize\epsffile{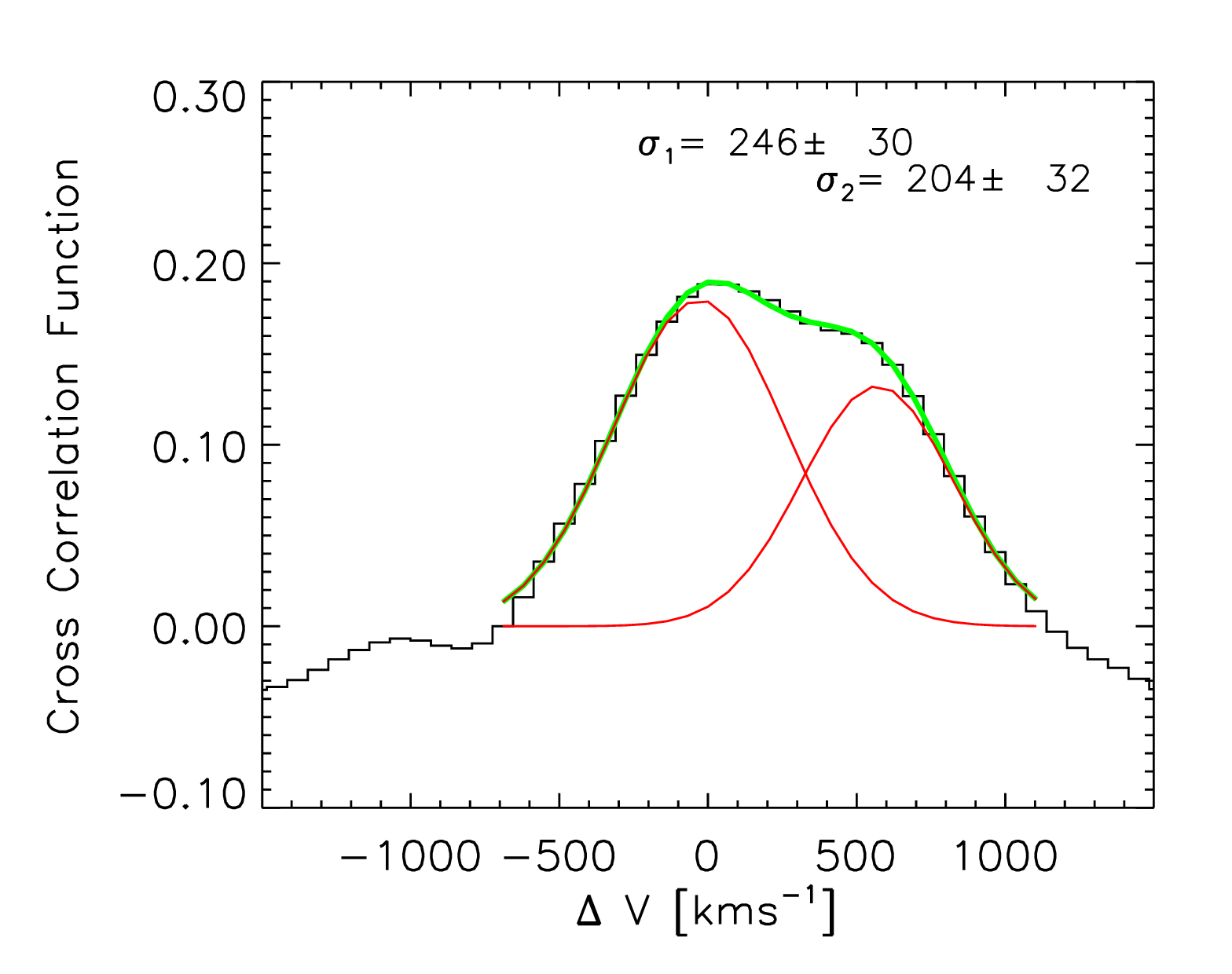}
 \epsfxsize=0.35\hsize\epsffile{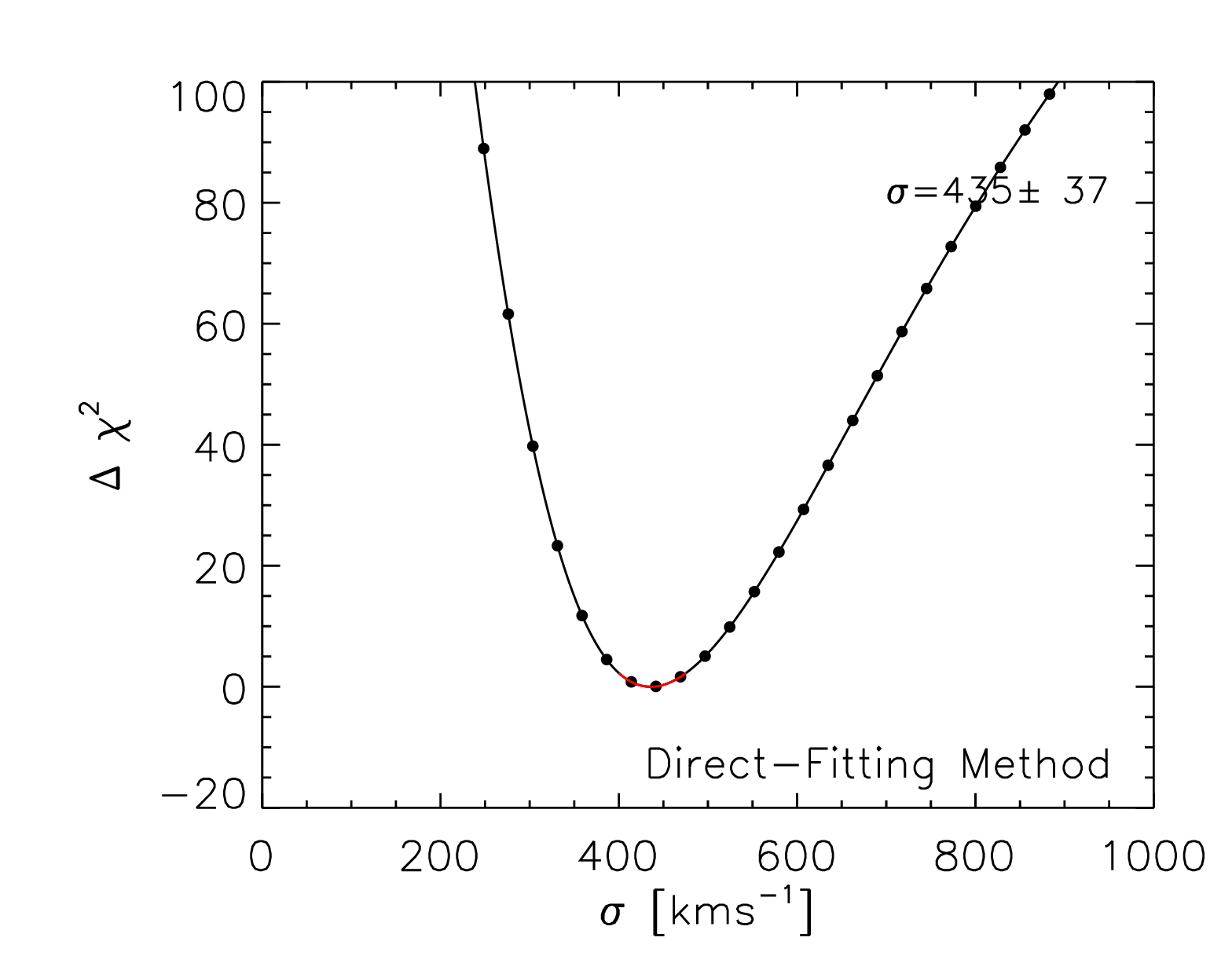}
 \epsfxsize=0.8\hsize\epsffile{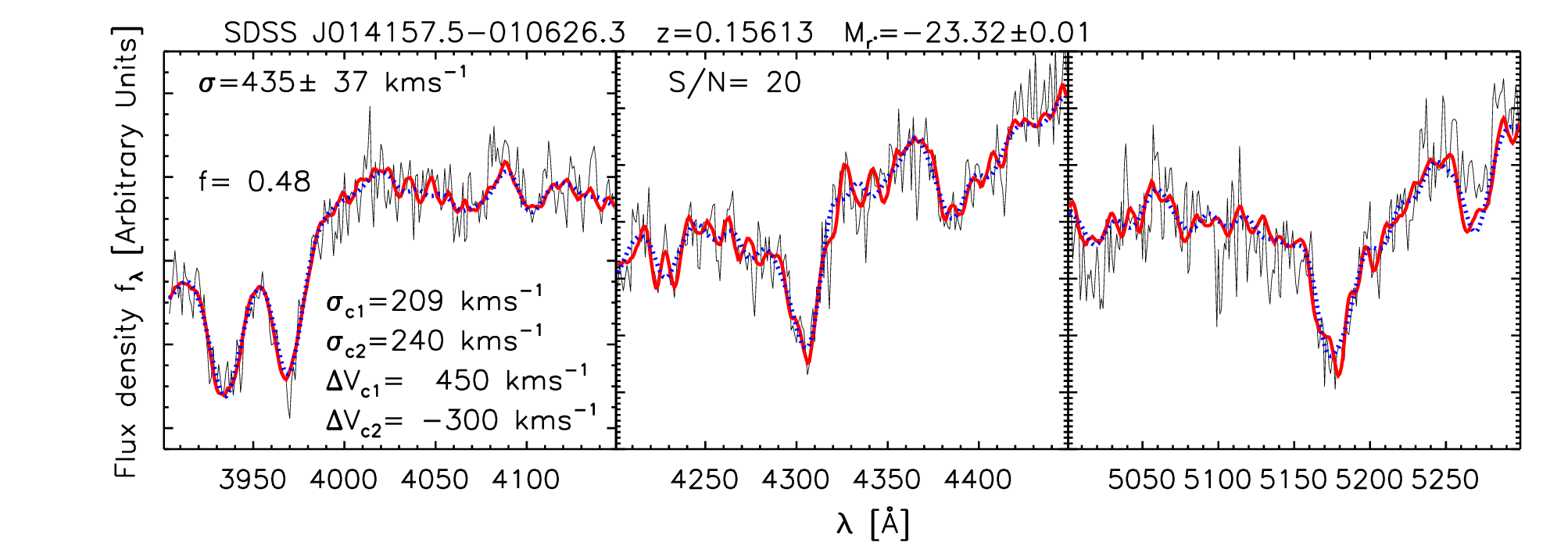}
 \caption{As in Figure~\ref{uniqueimgspec} but for objects classified as superpositions.}
 \label{doubleimgspec}
\end{figure*}

\setcounter{figure}{1}

\begin{figure*}
 \centering
 \epsfxsize=0.6\hsize\epsffile{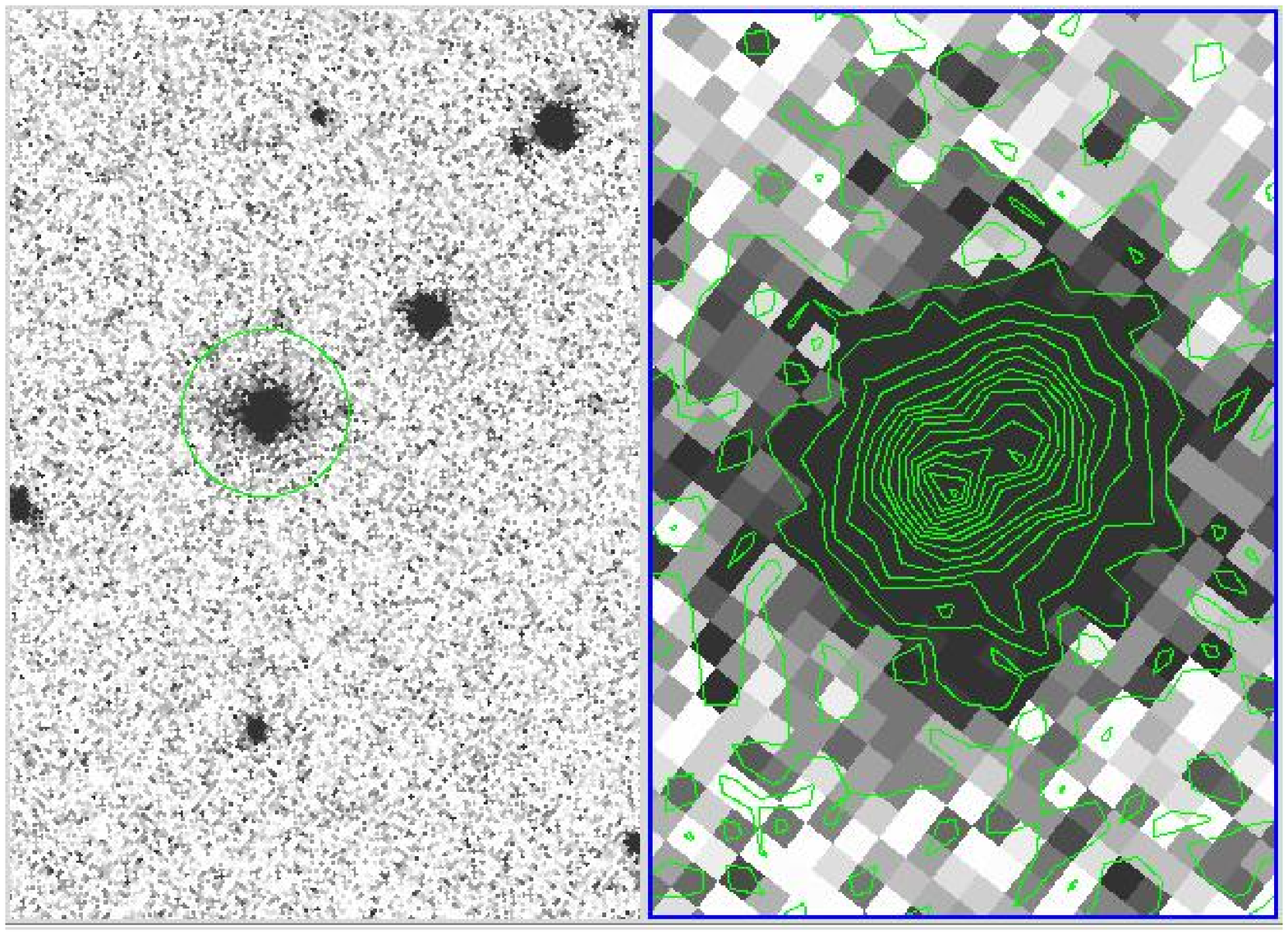}
 \epsfxsize=0.35\hsize\epsffile{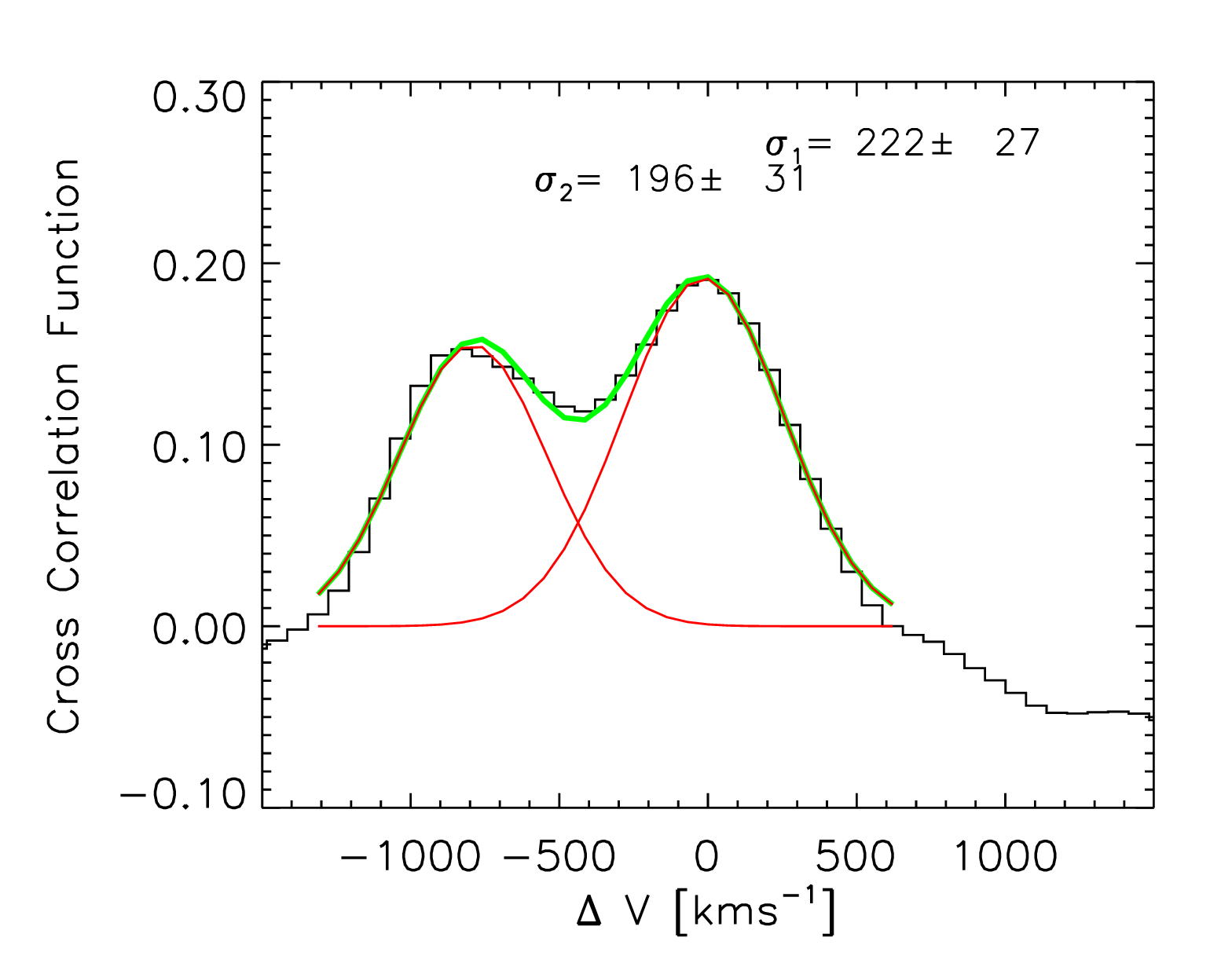}
 \epsfxsize=0.35\hsize\epsffile{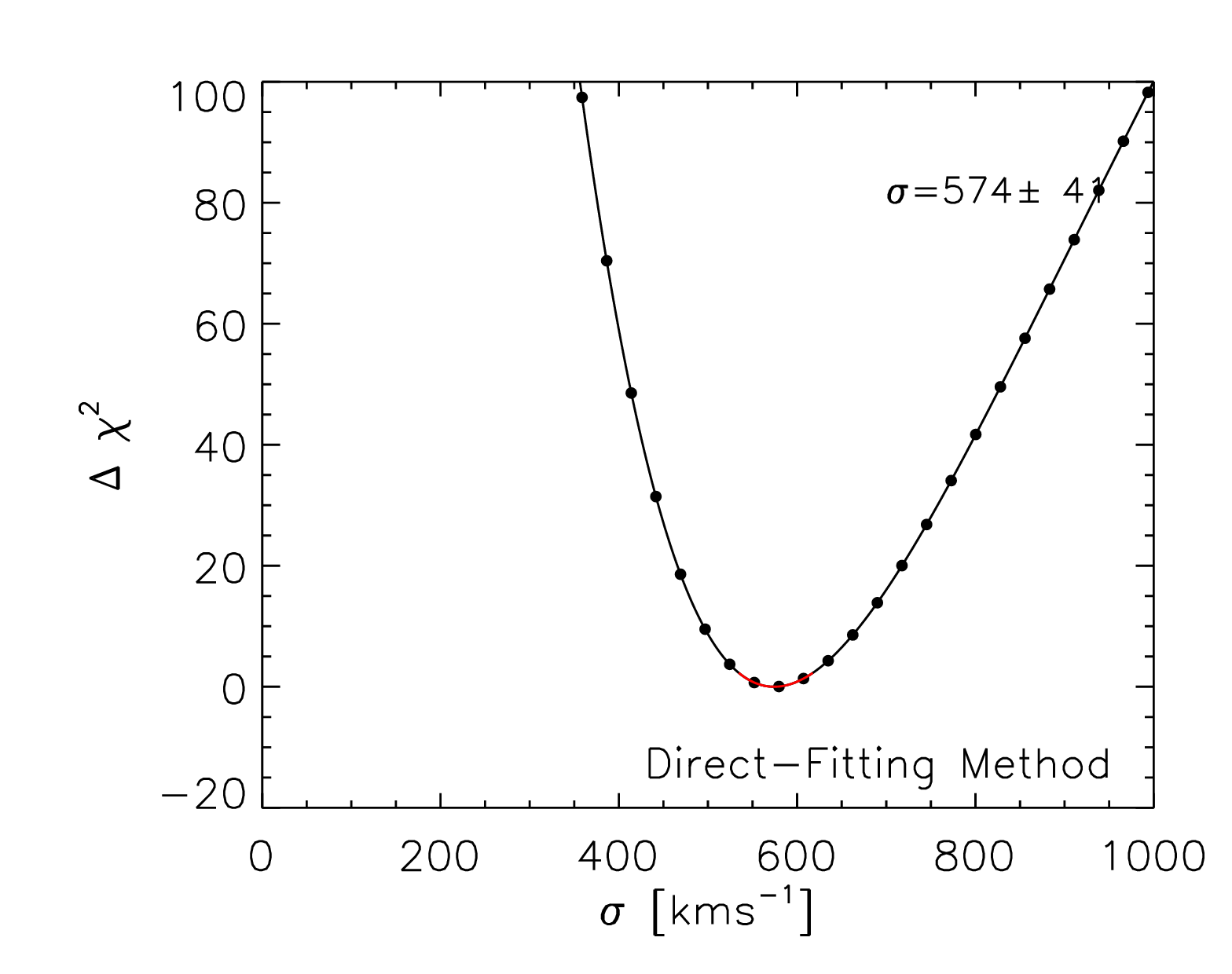}
 \epsfxsize=0.8\hsize\epsffile{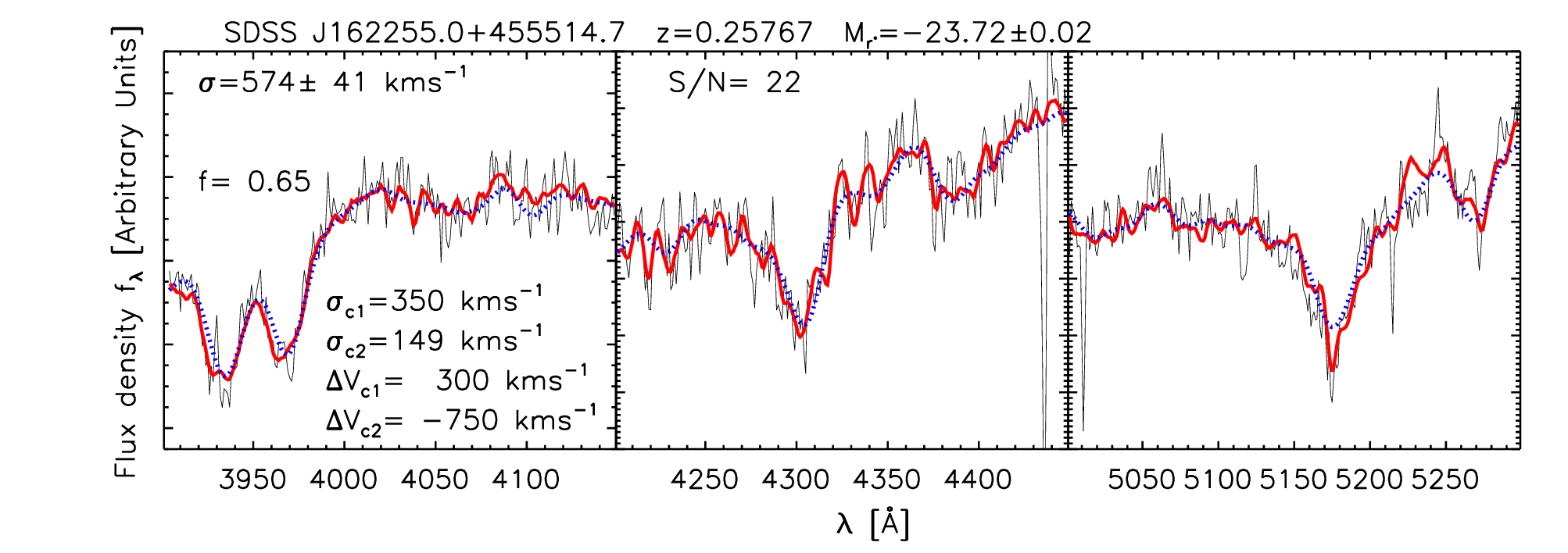}
 \caption{Continued.}
\end{figure*}

\setcounter{figure}{1}

\begin{figure*}
 \centering
 \epsfxsize=0.6\hsize\epsffile{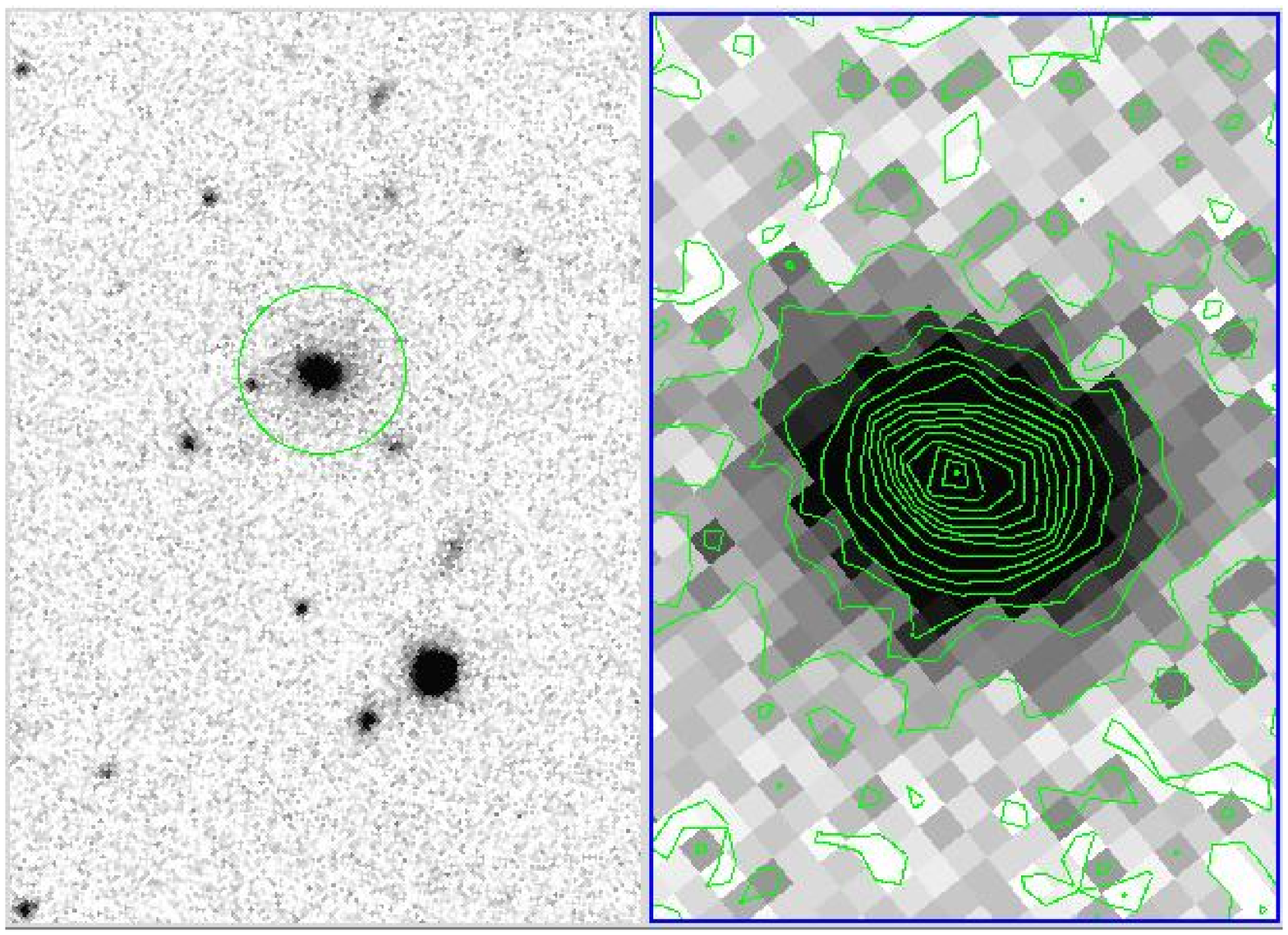}
 \epsfxsize=0.35\hsize\epsffile{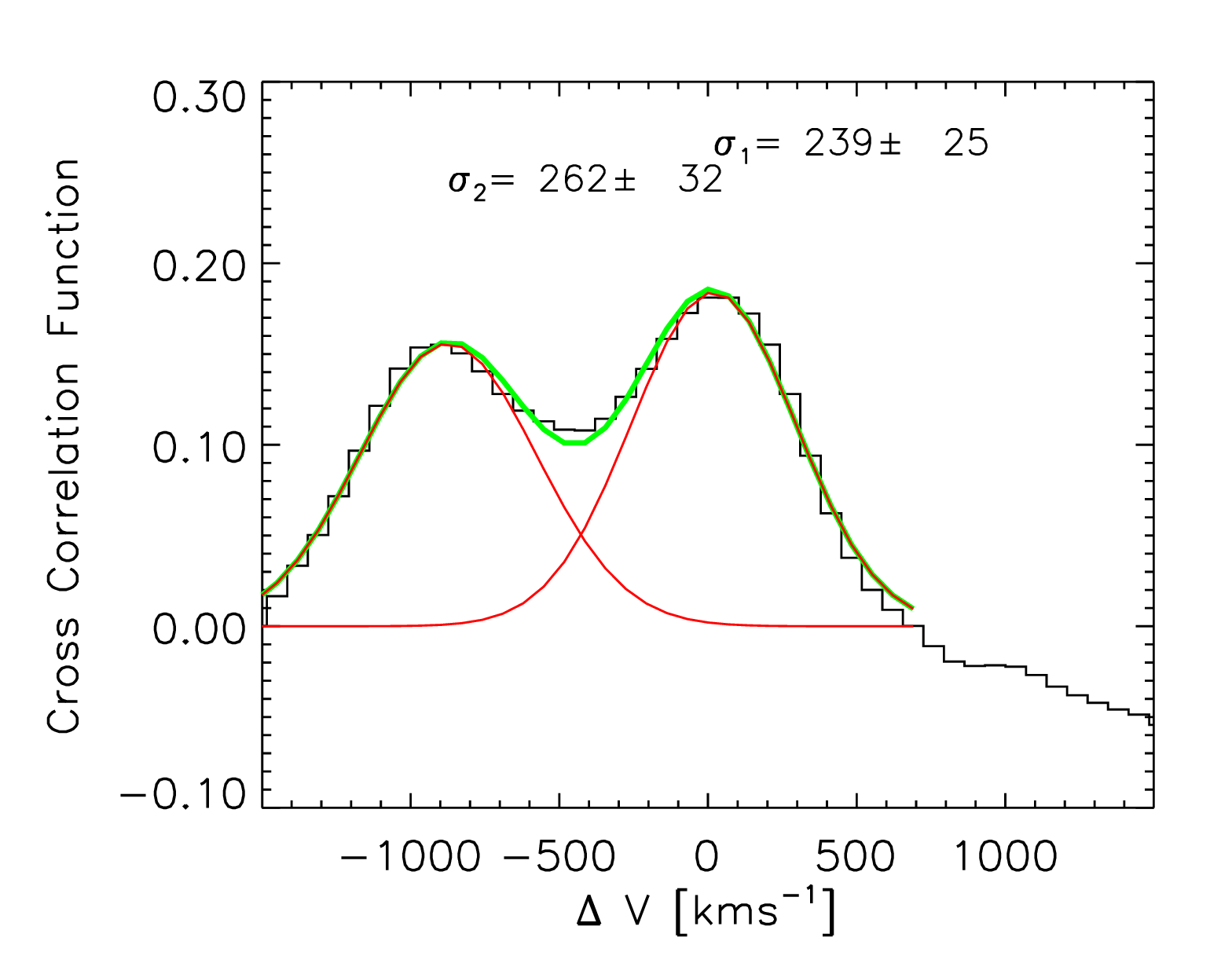}
 \epsfxsize=0.35\hsize\epsffile{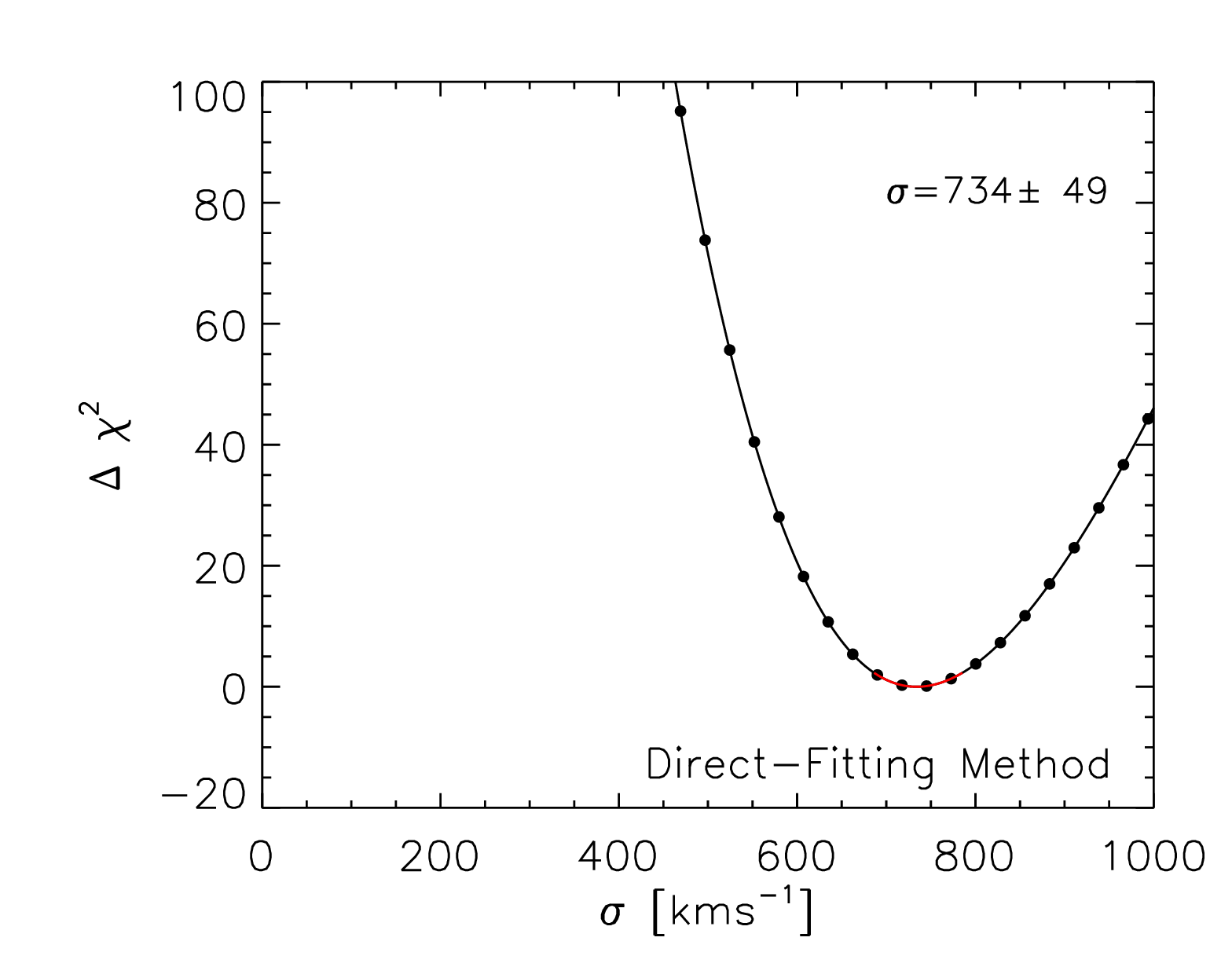}
 \epsfxsize=0.8\hsize\epsffile{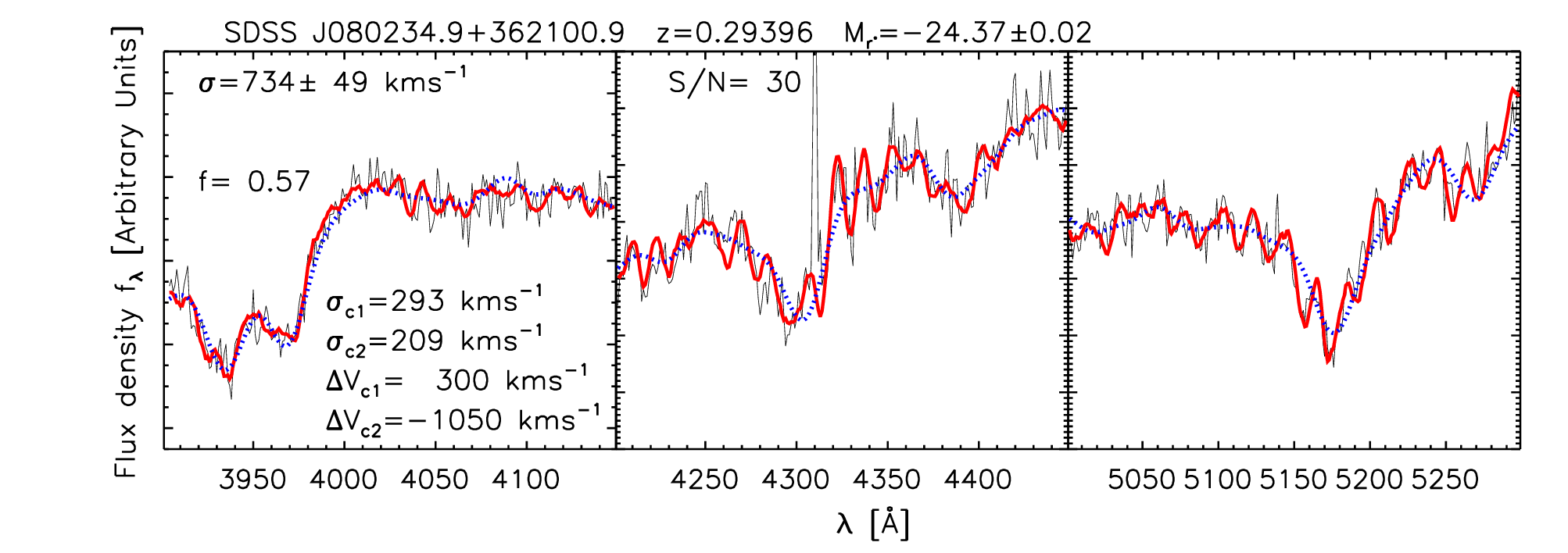}
 \caption{Continued.}
\end{figure*}

\end{document}